\documentclass[numbers]{elsarticle}
\usepackage{latexsym}
\usepackage{natbib}
\usepackage[latin2]{inputenc}
\usepackage{longtable}
\usepackage[hidelinks]{hyperref}

\makeatletter
\def\ps@pprintTitle{%
 \let\@oddhead\@empty
 \let\@evenhead\@empty
 \def\@oddfoot{\centerline{\thepage}}%
 \let\@evenfoot\@oddfoot}
\makeatother

\begin{document}

\title{The Advantage is at the Ladies: Brain Size Bias-Compensated Graph-Theoretical Parameters are Also Better in Women's Connectomes}

\author[p]{Balázs Szalkai}
\ead{szalkai@pitgroup.org}
\author[p]{Bálint Varga}
\ead{varga@pitgroup.org}
\author[p,u]{Vince Grolmusz\corref{cor1}}
\ead{grolmusz@pitgroup.org}
\cortext[cor1]{Corresponding author}
\address[p]{PIT Bioinformatics Group, Eötvös University, H-1117 Budapest, Hungary}
\address[u]{Uratim Ltd., H-1118 Budapest, Hungary}

\date{}

\begin{abstract}
In our previous study we have shown that the female connectomes have significantly better, deep graph-theoretical parameters, related to superior ``connectivity'', than the connectome of the males. Since the average female brain is smaller than the average male brain, one cannot rule out that the significant advantages are due to the size- and not to the sex-differences in the data. To filter out the possible brain-volume related artifacts, we have chosen 36 small male and 36 large female brains such that all the brains in the female set are larger than all the brains in the male set. For the sets, we have computed the corresponding braingraphs and computed numerous graph-theoretical parameters. We have found that (i) the small male brains lack the better connectivity advantages shown in our previous study for female brains in general; (ii) in numerous parameters, the connectomes computed from the large-brain females, still have the significant, deep connectivity advantages, demonstrated in our previous study.
\end{abstract}

\maketitle

\section{Introduction} 

While the neuronal-scale mapping of the connections of the whole human brain with more than 80 billion neurons is not possible today, a diffusion MRI-based workflow is available for mapping these connections with much less resolution \cite{Hagmann2012,Craddock2013a,McNab2013,Daducci2012}. The result of that workflow is the connectome, or the braingraph of the subject: the several hundred nodes of this graph correspond to distinct areas of the gray matter of the brain, and two nodes are connected by an edge if the workflow finds fibers of axons connecting the areas, corresponding to these two nodes. 

These connectomes describe tens of thousands of connections between distinct cerebral areas in a much more detailed manner than was possible before the era of diffusion MRI imaging. Additionally, the braingraphs make possible the quantitative analysis of the connections of the human brain. 

One natural question is finding the connections that are present in a majority of healthy subjects. In \cite{Szalkai2015a} we described the Budapest Reference Connectome Server \url{http://connectome.pitgroup.org} that generates and visualizes the consensus braingraph of healthy individuals according to selectable parameters. 

Another related question is the mapping of the individual variability of the connectomes in distinct areas of the gray matter. Using data from 392 healthy individuals, we have mapped the surprisingly different variability of the connections in distinct lobes and smaller cortical areas in \cite{Kerepesi2015c}. 

An interesting area is characterizing the significant differences in the brain-connections of distinct groups of subjects. Hundreds of publications appear describing differences of the connectomes of healthy and diseased individuals (e.g., \cite{Agosta2014,AlexanderBloch2014,Baker2014,Besson2014a,Bonilha2014}). 

Much fewer articles deal with sex differences of the structural properties of the connectomes. The authors of \cite{Sex2011} and \cite{Ingalhalikar2014b} have applied statistics for the numbers of edges connecting larger, fixed  anatomical areas of the cortex in men and women, and have found significant differences between the sexes in these numbers. The work \cite{Ingalhalikar2014b} analyzed the 95-vertex graphs of 949 subjects aged from  8 through 22 years on a publicly unavailable dataset. One of the main results of \cite{Ingalhalikar2014b} is showing that males have more intra-hemispheric edges while females have more inter-hemispheric edges.

Instead of simple edge-counting, we have applied much deeper -- even some NP-hard -- graph-theoretical algorithms for discovering sex differences between the connectomes in \cite{Szalkai2015}. The graph dataset examined in \cite{Szalkai2015} contained the data of 96 subjects of ages between 22 and 35 years, from the Human Connectome Project \cite{McNab2013}.  From the data of each subjects five graphs were computed with different resolutions and each graph with five different edge weights. The graph dataset is publicly available (without any registration) at \url{http://braingraph.org/download-pit-group-connectomes/} for independent verification and further analysis.  

We have found in \cite{Szalkai2015} that the female brain has such graph-theoretical properties that are associated with ``better connectivity'' in computer interconnection networks \cite{Tarjan1983a} and elsewhere. Namely, women's braingraphs have

\begin{itemize}
	\item[(i)] more edges;
	\item[(ii)] larger minimum bisection width (balanced minimum cut) within each hemisphere, even when normalized with the edge number;
	\item[(iii)] larger eigengap;
	\item[(iv)] more spanning forests;
	\item[(v)] larger minimum vertex cover
\end{itemize}

than the braingraphs of men. 
 
Property (i), the larger edge number, is a straightforward characteristic of a ``better connected'' graph. 
In computer interconnections networks, the quantity (ii) is a standard measure of the ``quality'' of the network \cite{Tarjan1983a}: the higher the width is the better the network. We note that the advantage remains valid even if we normalize the bisection width with the larger edge-number of the female connectomes! Quantity (iii) is related to the expander property of the graphs \cite{Hoory2006}: the larger the eigengap, the better expander is the graph. Good expander graphs have good intrinsic ``connectivity'' properties, such as a fast convergence to the stationary distribution of a random walk on an expander graph \cite{Hoory2006}.  

A minimum vertex cover is the smallest subset $S$ of the nodes of the graph such that each edge contains at least one vertex from $S$ (i.e., each edge is ``covered'' by an element of $S$).  The result in \cite{Szalkai2015} says that the edges in the braingraph of females need more vertices to cover than in the case of males.

It is known for a long time that on the average, the female brain weights less and is smaller than the brain of males \cite{Dekaban1978,Allen2002}. Clearly, this statistical difference may have implications for the diffusion MR imaging workflow, and, consequently, to the construction of the braingraphs from the imaging data. 

For example, larger brains have longer axonal fibers, and those longer fibers are more difficult to follow in the tractography phase of the data processing workflow: if we assume that there is a fixed error probability at every step of the tractography algorithm, then longer fibers will produce more errors, and, consequently, are much harder to follow and discover than short ones \cite{Girard2014,Jbabdi2011}. 

Therefore, it might happen that the statistically significant differences in the graph properties of the connectomes of different sexes are due to simple size differences. Clearly, any brain-size dependent ``artificial'' correction of the diffusion MRI data or the tractography results would meet well-founded criticism, and is not a realistic choice for data analysis.

The article \cite{Haenggi2014} states that the sex differences in the ratio of the intrahemispheric/interhemispheric connections in the  connectome are due to the differences in the brain size. In order to show this, they have selected 69 small brain subjects (55 females and 14 males) and 69 large brain subjects (14 females and 55 males), and after acquiring and analyzing their --- publicly unavailable --- data, they have found that the ratio of the intrahemispheric/interhemispheric streamlines (the authors of \cite{Haenggi2014} called it ``connectivity ratio'') is significantly larger in the small brain group than in the large brain group (0.218 vs. 0.201, p=0.005 from Table 1 of \cite{Haenggi2014}). 

We do not think that the results of \cite{Haenggi2014} are really decisive on the question since the large brain group contained mostly males and the small brain group mostly females.

\begin{table} [h!]
	\centering
	\includegraphics[width=4.7in]{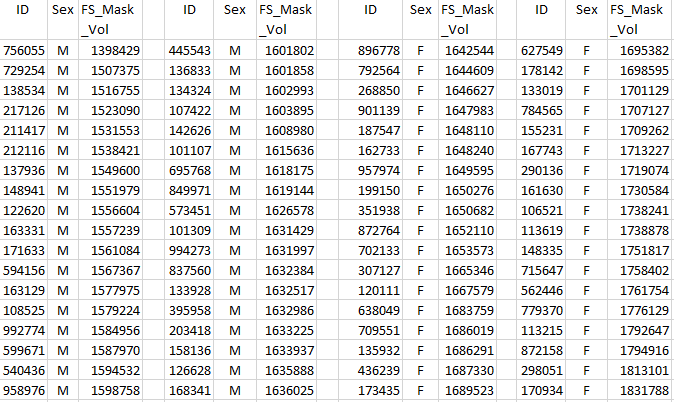}
	\caption{The list of subject-IDs, their sex and their brain volumes in the present study. The subjects are listed in the increasing order according to their brain volumes. The IDs refer to the Human Connectome Project's \cite{McNab2013} anonymized 500 Subjects Release. The corresponding braingraphs can be downloaded from the site \url{http://braingraph.org/download-pit-group-connectomes/}. The first six columns contain the data for 36 small-brain males, the last six columns the data of the 36 large-brain females. The columns with header FS\_Mask\_Vol contains the FreeSurfer-computed Brain Mask Volumes in mm$^3$ \cite{Reuter2012}.}
\end{table}

In our present work, we have computed braingraphs from the data of more than 400 subjects of the Human Connectome Project's \cite{McNab2013} anonymized 500 Subjects Release, and from this large set we were able to choose two rather unusual sets of subjects: 36 large-brain females and 36 small-brain males such that every single female brain is of larger volume than every single male brain in the set. Therefore, the sex-related differences in the connectomes in this set will be free from the size bias.

\section{Results and Discussion}

Here we consider the diffusion MRI data of small-brain males and large-brain females in order to compensate for the possible brain-size bias in the data acquisition and the data processing steps. 

We have chosen a set of 36 female and 36 male brains from the Human Connectome Project's \cite{McNab2013} anonymized 500 Subjects Release, such that each female brain in the set is larger than every male brain in the set (see Table 1). Next, as in \cite{Szalkai2015}, for every subject we have computed braingraphs of five different resolutions: 83, 129, 234, 463 and 1015 vertices, and for all graphs we have constructed five different edge weights: 

\begin{itemize}
	\item \texttt{Unweighted}: Every edge has the same, unit weight.
	\item \texttt{FiberN}: The number of fiber tracts that define the edge. 
	\item \texttt{FAMean}: The average of the fractional anisotropies \cite{Basser2011} of the fiber tracts, belonging to the edge. 
	\item \texttt{FiberLengthMean}: The average length of the fiber tracts belonging to the edge.
	\item \texttt{FiberNDivLength}: The count of the fiber tracts of the edge, divided by their average length. 
\end{itemize}

The most relevant weight function in our present study is the fractional anisotropy \texttt{FAMean}. This quantity, for each voxel, gives a measure of anisotropy with a real number between 0 and 1: 0, if the diffusion-ellipsoid in the voxel is a perfect sphere, and its value is getting closer to 1, if the ellipsoid has one large and two small axes, and it is 1 if the ellipsoid reduces to a line segment. 
More exactly,  One can assign a fractional anisotropy to fiber tracts by averaging the value for each voxels on the tract. The \texttt{FAMean} weight is the average of the fractional anisotropies, taken for all fiber tracts defining the graph-edge in question.

	\setlength\LTleft{-1.5cm}
	\setlength\LTright{-1.5cm}
	{\scriptsize
\begin{longtable}{r | rcccc}
	Nodes & Property & Group (F $|$ M) means & p (1st) & p (2nd) & p (corrected) \\ 
	83 & All\_AdjLMaxDivD\_FAMean & 1.353 $|$ 1.399 & 0.00195 & {\em 0.00003} & \textbf{0.00155} \\ 
	234 & All\_MinVertexCover\_FAMean & 52.243 $|$ 49.132 & 0.00298 & {\em 0.00008} & \textbf{0.00379} \\ 
	129 & All\_MinVertexCover\_FAMean & 29.504 $|$ 28.027 & 0.00867 & {\em 0.00073} & \textbf{0.03505} \\ 
	129 & All\_MinSpanningForest\_FAMean & 30.710 $|$ 27.945 & 0.005129 & {\em 0.00096} & \textbf{0.04507} \\ 
	234 & Left\_MinVertexCover\_FAMean & 26.154 $|$ 24.571 & 0.00062 & {\em 0.00113} & 0.05192 \\ 
	83 & All\_MaxMatching\_FAMean & 18.693 $|$ 17.688 & 0.00408 & {\em 0.00120} & 0.05407 \\ 
	83 & Right\_LogSpanningForestN\_FAMean & 51.672 $|$ 47.347 & 0.00422 & {\em 0.00197} & 0.08681 \\ 
	83 & All\_MaxFracMatching\_FAMean & 18.744 $|$ 17.795 & 0.00283 & {\em 0.00199} & 0.08575 \\ 
	83 & All\_MinVertexCover\_FAMean & 18.744 $|$ 17.795 & 0.00283 & {\em 0.00199} & 0.08376 \\ 
	83 & All\_LogSpanningForestN\_FAMean & 110.058 $|$ 101.687 & 0.00287 & {\em 0.00365} & 0.14977 \\ 
	83 & All\_MinSpanningForest\_FAMean & 20.162 $|$ 18.323 & 0.00538 & {\em 0.00428} & 0.17106 \\ 
	83 & Right\_Sum\_FAMean & 103.256 $|$ 94.541 & 0.00240 & {\em 0.00567} & 0.22122 \\ 
	129 & Left\_MinVertexCover\_FAMean & 14.647 $|$ 13.925 & 0.00217 & {\em 0.00585} & 0.22215 \\ 
	234 & All\_LogSpanningForestN\_FAMean & 334.237 $|$ 308.236 & 0.00428 & {\em 0.00593} & 0.21951 \\ 
	83 & Left\_LogSpanningForestN\_FAMean & 53.522 $|$ 48.713 & 0.00990 & {\em 0.00746} & 0.26874 \\ 
	129 & All\_LogSpanningForestN\_FAMean & 192.234 $|$ 179.953 & 0.00714 & {\em 0.00833} & 0.29154 \\ 
	234 & All\_Sum\_FAMean & 683.263 $|$ 630.432  & 0.00295 & {\em 0.01170} & 0.39796 \\ 
	129 & Left\_LogSpanningForestN\_FAMean & 95.757 $|$ 89.284 & 0.00401 & {\em 0.01603} & 0.52905 \\ 
	234 & Left\_LogSpanningForestN\_FAMean & 165.628 $|$ 152.443  & 0.00043 & {\em 0.01635} & 0.52333 \\ 
	83 & All\_Sum\_FAMean & 217.684  $|$ 201.464  & 0.00510 & {\em 0.01788} & 0.55437 \\ 
	83 & Left\_MaxFracMatching\_FAMean & 9.242 $|$ 8.803  & 0.00192 & {\em 0.02533} & 0.76004 \\ 
	83 & Left\_MinVertexCover\_FAMean & 9.242 $|$ 8.803  & 0.00192 & {\em 0.02533} & 0.73470 \\ 
	83 & Left\_MaxMatching\_FAMean & 9.187 $|$ 8.742  & 0.00408 & {\em 0.02558} & 0.71621 \\ 
	129 & All\_Sum\_FAMean & 386.878 $|$ 360.531  & 0.00566 & {\em 0.02624} & 0.70860 \\ 
	234 & Left\_Sum\_FAMean & 338.278 $|$ 313.589  & 0.00264 & {\em 0.03474} & 0.90314 \\ 
	234 & All\_MinVertexCover\_FiberLengthMean & 5345.241 $|$ 4942.425  & 0.00624 & {\em 0.03583} & 0.89570 \\ 
	129 & All\_MinSpanningForest\_FiberLengthMean & 1671.711 $|$ 1639.914  & 0.00419 & {\em 0.03734} & 0.89622 \\ 
	129 & Left\_Sum\_FAMean & 193.159 $|$ 179.854  & 0.00640 & {\em 0.04110} & 0.94530 \\ 
	83 & Right\_Sum\_FiberLengthMean & 7597.945 $|$ 6933.002  & 0.00616 & 0.05527 & 1.21602 \\ 
	83 & Left\_MinSpanningForest\_FAMean & 9.767 $|$ 9.066  & 0.00644 & 0.05603 & 1.17663 \\ 
	234 & All\_MaxMatching\_FiberLengthMean & 5329.953 $|$ 4998.430  & 0.00501 & 0.05622 & 1.12431 \\ 
	234 & Left\_AdjLMaxDivD\_FAMean & 1.552 $|$ 1.597  & 0.00188 & 0.06919 & 1.31463 \\ 
	83 & All\_Sum\_FiberLengthMean & 16777.596 $|$ 15541.031  & 0.00895 & 0.07094 & 1.27698 \\ 
	234 & All\_MaxFracMatching\_FiberLengthMean & 5341.519 $|$ 5025.310  & 0.00595 & 0.07252 & 1.23279 \\ 
\caption{Statistical analysis of the differences in graph-theoretical parameters of connectomes computed from 36 small-brain male and 36 large-brain  female cerebral MRIs. The first column gives the number of vertices. The second column gives the graph-parameters computed: Each parameter-name is separated by two ``$\_$'' symbols into three segments: The first segment describes the hemisphere or the whole connectome using descriptors Left, Right or All. The second segment gives the graph-parameter computed (defined in the ``Methods'' section, e.g.,\texttt{MaxMatching}), and the third segment specifies the weight function applied, the choices are \texttt{Unweighted, FiberN, FAMean, FiberLengthMean, FiberNDivLength}.
The third and the fourth columns contain the average values for the female and the male groups, respectively. The fifth column describes the p-values of the first round, the sixth column in the second round, and the seventh column the Holm-Bonferroni corrections for the p-values. 
  With p=0.05 {\em all} the first four rows describe significantly different graph theoretical properties between sexes. One-by-one, each row with italic values in column 6 describe differences between sexes, with significance p=0.05. For the details we refer to the section ``Statistical analysis''.} 
\end{longtable}

}

\subsection*{The lack of both significant and non-significant advantages of males with small brains}

Suppose that the statistically significant differences of the graph-theoretical parameters, describing better connectivity of the female braingraphs in \cite{Szalkai2015}, are due solely to the brain volume differences and not to the sex differences of the subjects.  Then in our subject-sets of small-brain men and large-brain women the same, significant differences in the graph-theoretical parameters should have been observed showing the advantage of the small-brain males. 

Surprisingly, this is not true, with one single exception: 

As one can observe in the first row of Table 2, \texttt{All\_AdjLMaxDivD\_FAMean} (the largest eigenvalue of the generalized adjacency matrix, divided by the average (generalized) degree, computed with the \texttt{FAMean} weight function)  is significantly larger for men than for women in the 83-vertex resolution. All the other parameters that were significant statistically (denoted by an italic font in the sixth column) are still larger for the female group, showing better connections in that group. 

This means that the lower cerebral volume will not imply better connectivity, therefore the results of \cite{Szalkai2015} are due to sex differences and not size differences or other artifacts.

\subsection*{FAMean-weighted significant connectivity advantages of females}

We need to remark that almost all graph-parameters, implying better connectivity for the connectomes of women in Table 2, were weighted by  \texttt{FAMean}. We think that fiber tracts with high \texttt{FAMean} values were tracked very reliably and were able to produce statistically significant results. 

More exactly, we have found (in Table 2) the following parameters significantly differing in both tests: 

\texttt{MinVertexCover} for the whole brain, and also just for the left hemisphere in several resolutions: this quantity gives the minimum number of vertices that is needed to cover all edges in the graph. We note that the meaning of this NP-hard quantity and also its computing needs much deeper tools than the numerous edge-counting statistics published elsewhere.

\texttt{Sum} describes the number of edges in the graph; we have significant differences within the left and the right hemispheres, as well as in the whole connectome.
The number of spanning forests also give significant advantages for the female connectomes. Similarly, the maximum matching and the maximum fractional matching is also significantly larger in female connectomes for several resolutions. 

In the ANOVA round 1, when only the data from group 0 were analyzed, we have found lots of significant-looking results showing the better connectivity of the female brains even in our large-brain set with other weight functions as well. For example, in the Appendix, in the 463 node resolution,  the minimum number of the vertices that are needed to cover all the edges in the whole braingraph is larger in female connectomes than in male ones even with the \texttt{FiberN} weight function, with $p=0.017$. Or, in the same 463-node resolution, the minimum balanced bisection width, normalized with the number of edges in the right hemisphere is larger for women than in men with $p=0.05$ in the unweighted graph. We say that these are ``significant-looking'' results since our strict analysis in the holdout set did not find all of these to be significant (see the ``Statistical Analysis'' section for details).

In summary, with the \texttt{FAMean} weight function several (but not all) differences we have found between the graph parameters of male and female connectomes remained valid for the small-brain men - large-brain women datasets. Additionally, with numerous other weight functions the advantage of the female connectomes in connectivity related parameters is shown in the Appendix.

\section{Methods}

The data source is the  Human Connectome Project's \cite{McNab2013} anonymized 500 Subjects Release. The workflow that produces the braingraphs or connectomes are detailed in \cite{Szalkai2015} and in \cite{Szalkai2015a}: the Connectome Mapper Toolkit \cite{Daducci2012} (\url{http://cmtk.org}) was used for segmentation, partitioning, tractography and for the graph construction. For partitioning, FreeSurfer was applied with the Desikan-Killiany anatomical atlas that produced 83, 129, 234, 463 and 1015 regions of interests. Tractography was performed  with randomized seeding by the Connectome Mapper Toolkit \cite{Daducci2012}, applying the deterministic streamline method with the MRtrix processing tool \cite{Tournier2012}. We have computed the graphs from more than 400 diffusion MR images.

\subsection*{Choosing two sets of the same cardinality: large female brains and small male brains} 

The selection of large female and small male brains were done using the following mathematical scheme: 

There is such a brain size B that an equal number of men have smaller brains than B as the women who have larger brains than B. This is true because each person has a different brain size, and when we increase B from the minimum possible brain size to the maximum possible brain size, at each step we either encounter a man (in this case the number of men with smaller brains than B increases by 1), or we encounter a woman (in this case the number of women with larger brains than B decreases by 1). Since at the beginning the number of small-brain men was 0 and the number of large-brain women was $N_W$ (the number of women in the study), and at the end the number of small-brain men will be $N_M$ (the number of men in the study) and the number of large-brain women will be 0, this means that at some point the two numbers will be equal because, at each step, both change by 1 in the proper direction. This is a well-defined interval between two consecutive brain sizes. We looked for this division point B, and then considered only the men with smaller brains and the women with larger brains.

This way we were able to select 36 female and 36 male brains, such that all the female brains have larger volumes than all the male brains in the set.

\subsection*{Graph parameters and their descriptions}

The {\em generalized adjacency matrix} is an $n\times n$ matrix, where $n$ denotes the number of vertices of the graph. Its rows and columns correspond to the nodes of the graph, and the element in the intersection of row $i$ and column $j$, $a_{ij}$ is zero if the $i^{th}$ and the $j^{th}$ vertices are not connected by an edge, and $a_{ij}$ is the weight of the edge, connecting the $i^{th}$ and the $j^{th}$ vertices otherwise.

The degree of a node is the number of the edges, incident to the vertex. The generalized degree of a vertex is the sum of the weights of the edges, incident to the vertex.

The following graph-parameters were computed for the graphs of different resolutions and weights:

\begin{itemize}
	\item Number of edges (\texttt{Sum}). The sum of the weights of the edges. If the graph is unweighted, then it is equal to the number of edges in the graph.
	
	\item Normalized largest eigenvalue (\texttt{AdjLMaxDivD}): The largest eigenvalue of the generalized adjacency matrix, divided by the average (generalized) degree. 	
	
	\item Eigengap of the transition matrix (\texttt{PGEigengap}): The transition matrix $P_G$ can be constructed after dividing the rows of the generalized adjacency matrix by the generalized degree of the vertex, corresponding to the row. Since the (generalized) degree of any vertex is equal to the sum of its row in the generalized adjacency matrix, the sum of any row of $P_G$ is 1. If the weights are non-negative, then the rows of $P_G$ define a probability distribution, which corresponds to the transition probabilities in a random walk.  The eigengap of matrix $P_G$ is the difference between its largest and the second largest eigenvalues. The eigengap is closely related to the expander property of the graph: the larger the gap, the better expander is the graph \cite{Hoory2006}.
	
	\item Hoffman's bound (\texttt{HoffmanBound}) is defined by $$1 + \frac{\lambda_{max}}{|\lambda_{min}|},$$ where $\lambda_{max}$ and $\lambda_{min}$ denote the largest and smallest eigenvalues of the adjacency matrix. It bounds the chromatic number of the graph from below. 
		
	\item Logarithm of the number of spanning forests (\texttt{LogAbsSpanningForestN}): The famous matrix-tree theorem of Kirchoff \cite{Kirchoff1847,Chung1997} computes the number of the spanning trees in a connected graph from the spectrum of its Laplacian matrix. Heuristically, more ``connected'' graphs have more spanning trees, since the addition of a new edge to the graph may give rise to the number of the spanning trees. For non-connected graphs, the number of spanning forests equals the product of the numbers of the spanning trees of the components of the graph. The quantity \texttt{LogAbsSpanningForestN} is defined as 
	the logarithm of the number of spanning forests in the unweighted case; and in the weighted case it equals to the sum of the logarithms of the weights of the spanning trees in the forests.	Note that this value can be negative if all the weights are small.
	
	\item Minimum bisection width, or the balanced minimum cut, divided by the number of edges (\texttt{MinCutBalDivSum}): Suppose we want to partition the graph into two sets whose size may differ by at most 1, in a way that the the number (or the sum of the weights) of the edges, crossing the cut, is minimal. For the whole braingraph, one would expect that this minimum cut corresponds to the edges in the {\it corpus callosum} between the two hemispheres of the brain. Indeed, our results show exactly this. Therefore, this quantity is more interesting when computed only for the left- or the right hemisphere, and not for the whole brain.
	
	\item Minimum cost spanning tree (\texttt{MinSpanningForest}), computed with the algorithm of Kruskal. 
	
	\item Minimum vertex cover (\texttt{MinVertexCoverBinary}): is the size of the minimum set of vertices selected in a way that each edge is incident to at least one of the selected vertices. We have computed this NP-hard graph-parameter only for unweighted graphs by an integer-program solver named SCIP (\url{http://scip.zib.de}), \cite{Achterberg2008, Achterberg2009}, which provided exact solutions.
	
	\item Minimum weighted vertex cover (\texttt{MinVertexCover}): We assign a fractional weight to each vertex such that, for each edge, the sum of the weights of its two endpoints is greater or equal to 1, then we minimize the sum of all weights for all vertices. This is a relaxation of the vertex-cover problem above \cite{Hochbaum1982}, and can be computed by a linear programming approach.
	
 
  \item Maximum matching (\texttt{MaxMatching}): A matching is a set of edges that do not share any vertices; or, equivalently: each vertex covered by the matching edges are covered by exactly one edge from the matching. A maximum matching is the matching in a graph containing the largest number of edges. A maximum matching in a weighted graph is the matching with the maximum sum of weights taken on its edges.
 
 \item Maximum fractional matching (\texttt{MaxFracMatching}): is the linear-programming relaxation of the maximum matching problem. In the unweighted case, we are searching for non-negative values $x(e)$ for each edge $e$ in the graph, such that for each vertex $v$ in the graph, the sum of $x(e)$-s for the edges that are incident to $v$ is at most 1. The maximum of the sums of $\sum_ex(e)$ is the maximum fractional matching for a graph. For the weighted version with weight function $w$, instead of $\sum_ex(e)$, $\sum_ex(e)w(e)$ needs to be maximized.

 \item (\texttt{OutBasalGanglia, OutBrainstem, OutFrontal, OutInsula, OutLimbic, OutOccipital, OutParietal, OutTemporal, OutThalamus}) These quantities give the sum of the weights of the edges, crossing the border of the cerebral lobes noted.
	
\end{itemize}

All the parameters described above were computed for the graphs made of the left and the right hemispheres and also for the whole connectome, and for all the resolutions and with all the 5 weight functions (with the following exceptions: \texttt{MinVertexCoverBinary} and \texttt{MaxMatching}  was computed only for the unweighted case, and the \texttt{MinSpanningTree} was not computed for the unweighted case). 
The results for each individual graph are made available as a large Excel file at the site \url{http://uratim.com/big_table_sbmbbw.zip}.
	
\subsection*{A note on the syntax of the results} 
Each parameter-name is separated by two ``$\_$'' symbols into three segments (e.g., \texttt{All\_HoffmanBound\_FAMean}): The first segment describes the hemisphere or the whole connectome using descriptors Left, Right or All. The second segment gives the graph-parameter computed (defined in the ``Methods'' section, e.g.,\texttt{HoffmanBound}), and the third segment specifies the weight function applied, the choices are \texttt{Unweighted, FiberN, FAMean, FiberLengthMean, FiberNDivLength}. The weight functions are defined in the ``Results and discussion'' section.
	
\subsection*{Statistical analysis}

The statistical analysis of the sex differences in the graph-theoretical parameters were done similarly as in \cite{Szalkai2015}: 

The subjects were divided into two sets: set 0 and set 1, denoted in the first column of the large, detailed result file at \url{http://uratim.com/big_table_sbmbbw.zip}. The selection was done by the parity of the digits of the ID of the subjects: if the sum of the digits of the ID number of the subject was even, the ID was assigned to group 0, and if it was odd, then to group 1. Group 0 was used for hypothesis-making while group 1 was the holdout set to verify hypotheses. 

We applied the statistical null hypothesis \cite{Hoel1984} that the graph parameters do not differ between the male and the female groups. A small $p$ value shows that the null-hypothesis is most probably false, i.e., the graph parameters significantly differ between the sexes. 

We have used ANOVA (Analysis of variance) \cite{Wonnacott1972} to assign p-values for all parameters in each hemisphere and each resolution and each weight-assignment for data, originated from group 0. 

We selected those parameters after the first ANOVA application where the p-values were less than 1\%. These selected parameters were analyzed with a second ANOVA application for the holdout group 1. Next, the resulting p-values were adjusted with the Holm-Bonferroni correction method \cite{Holm1979} with a significance level of 5\%. The detailed results with the male and female average values of the parameters with the p-values of the first ANOVA are given in the Appendix, grouped by the resolution of the graph. The results of the second ANOVA and the Holm-Bonferroni corrections are given on Table 2.

In Table 2 those Holm-Bonferroni corrected p-values were highlighted in bold that all differs significantly between the male and the female groups, with a level of significance of 5\%.

\section{Conclusions}

We have shown by analyzing the connectomes of 36 small-brain men and 36 large-brain women that the advantage of the female connectomes in numerous graph-connectivity related, deep graph theoretical parameters, are due to the sex differences, and not for the size differences.

\section*{Data availability:} The raw and the pre-processed MRI data are available at the Human Connectome Project's website:
\url{http://www.humanconnectome.org/documentation/S500} \cite{McNab2013}.  Unlike numerous braingraph-related articles, our graphs that we assembled in the present work can be downloaded from the site \url{http://braingraph.org/download-pit-group-connectomes/}. The results for each individual graph are made available as a large Excel file at the site \url{http://uratim.com/big_table_sbmbbw.zip}.

\section*{Acknowledgments}
Data were provided in part by the Human Connectome Project, WU-Minn Consortium (Principal Investigators: David Van Essen and Kamil Ugurbil; 1U54MH091657) funded by the 16 NIH Institutes and Centers that support the NIH Blueprint for Neuroscience Research; and by the McDonnell Center for Systems Neuroscience at Washington University.



\section{Appendix}

In this Appendix we give the graph-theoretic parameters computed for the 83, 129, 234, 463 and 1015-vertex graphs. The table contains their arithmetic means in the male and female groups, and the corresponding p-values for group 0 (see the ``Statistical analysis'' subsection). The graph-parameters and the syntax of the data are defined in the ``Methods'' section. Significant differences ($p<0.01$) are denoted with an asterisk in the last column.

\small{
	\subsection{83 nodes, round 1}
	\label{Table_Round1_33}
	\begin{longtable}{l | cccc}
		Property & Female & Male & p-value &  \\ 
		All\_AdjLMaxDivD\_FAMean & 1.35697 & 1.39840 & 0.00195 & $*$ \\ 
		All\_AdjLMaxDivD\_FiberLengthMean & 1.47297 & 1.42258 & 0.07853 \\ 
		All\_AdjLMaxDivD\_FiberN & 2.06602 & 2.13826 & 0.22061 \\ 
		All\_AdjLMaxDivD\_FiberNDivLength & 1.85001 & 1.87343 & 0.56770 \\ 
		All\_AdjLMaxDivD\_Unweighted & 1.26011 & 1.26981 & 0.36427 \\ 
		All\_HoffmanBound\_FAMean & 4.21756 & 4.11377 & 0.10446 \\ 
		All\_HoffmanBound\_FiberLengthMean & 3.18204 & 3.21898 & 0.60378 \\ 
		All\_HoffmanBound\_FiberN & 2.61190 & 2.58253 & 0.58132 \\ 
		All\_HoffmanBound\_FiberNDivLength & 2.43437 & 2.49011 & 0.41972 \\ 
		All\_HoffmanBound\_Unweighted & 4.48639 & 4.39983 & 0.21541 \\ 
		All\_LeftRatio\_FAMean & 0.96460 & 0.95909 & 0.72266 \\ 
		All\_LeftRatio\_FiberLengthMean & 1.01590 & 1.01323 & 0.89446 \\ 
		All\_LeftRatio\_FiberN & 0.99308 & 0.99250 & 0.96520 \\ 
		All\_LeftRatio\_FiberNDivLength & 0.98904 & 0.98939 & 0.97426 \\ 
		All\_LeftRatio\_Unweighted & 0.99523 & 0.99065 & 0.63900 \\ 
		All\_LogSpanningForestN\_FAMean & 107.94151 & 97.08760 & 0.00287 & $*$ \\ 
		All\_LogSpanningForestN\_FiberLengthMean & 455.16733 & 445.91974 & 0.02158 \\ 
		All\_LogSpanningForestN\_FiberN & 395.24633 & 390.93348 & 0.03833 \\ 
		All\_LogSpanningForestN\_FiberNDivLength & 146.08409 & 145.33936 & 0.72755 \\ 
		All\_LogSpanningForestN\_Unweighted & 190.04825 & 186.55742 & 0.04411 \\ 
		All\_MaxFracMatching\_FAMean & 18.60828 & 17.28793 & 0.00283 & $*$ \\ 
		All\_MaxFracMatching\_FiberLengthMean & 2036.50193 & 1821.53984 & 0.01830 \\ 
		All\_MaxFracMatching\_FiberN & 2347.16667 & 2391.85714 & 0.49338 \\ 
		All\_MaxFracMatching\_FiberNDivLength & 108.48762 & 109.71109 & 0.73742 \\ 
		All\_MaxFracMatching\_Unweighted & 40.90000 & 40.85714 & 0.78684 \\ 
		All\_MaxMatching\_FAMean & 18.54438 & 17.26157 & 0.00408 & $*$ \\ 
		All\_MaxMatching\_FiberLengthMean & 2041.39713 & 1823.19540 & 0.02283 \\ 
		All\_MaxMatching\_FiberN & 2332.46667 & 2394.14286 & 0.32065 \\ 
		All\_MaxMatching\_FiberNDivLength & 107.64318 & 109.83960 & 0.53256 \\ 
		All\_MaxMatching\_Unweighted & 40.60000 & 40.57143 & 0.88134 \\ 
		All\_MinCutBalDivSum\_FAMean & 0.03843 & 0.04534 & 0.09680 \\ 
		All\_MinCutBalDivSum\_FiberLengthMean & 0.02920 & 0.03626 & 0.15883 \\ 
		All\_MinCutBalDivSum\_FiberN & 0.02789 & 0.02766 & 0.94947 \\ 
		All\_MinCutBalDivSum\_FiberNDivLength & 0.03111 & 0.03071 & 0.90781 \\ 
		All\_MinCutBalDivSum\_Unweighted & 0.03567 & 0.04166 & 0.10640 \\ 
		All\_MinSpanningForest\_FAMean & 19.76563 & 17.81751 & 0.00538 & $*$ \\ 
		All\_MinSpanningForest\_FiberLengthMean & 1102.02369 & 1084.66325 & 0.19564 \\ 
		All\_MinSpanningForest\_FiberN & 99.00000 & 104.00000 & 0.07431 \\ 
		All\_MinSpanningForest\_FiberNDivLength & 3.47983 & 3.92020 & 0.01709 \\ 
		All\_MinVertexCoverBinary\_Unweighted & 59.33333 & 58.85714 & 0.41628 \\ 
		All\_MinVertexCover\_FAMean & 18.60828 & 17.28793 & 0.00283 & $*$ \\ 
		All\_MinVertexCover\_FiberLengthMean & 2036.50193 & 1821.53984 & 0.01830 \\ 
		All\_MinVertexCover\_FiberN & 2347.16667 & 2391.85714 & 0.49338 \\ 
		All\_MinVertexCover\_FiberNDivLength & 108.48762 & 109.71109 & 0.73742 \\ 
		All\_MinVertexCover\_Unweighted & 40.90000 & 40.85714 & 0.78684 \\ 
		All\_PGEigengap\_FAMean & 0.05014 & 0.05685 & 0.18480 \\ 
		All\_PGEigengap\_FiberLengthMean & 0.03839 & 0.04616 & 0.19918 \\ 
		All\_PGEigengap\_FiberN & 0.02904 & 0.02801 & 0.67953 \\ 
		All\_PGEigengap\_FiberNDivLength & 0.03133 & 0.03050 & 0.73932 \\ 
		All\_PGEigengap\_Unweighted & 0.04661 & 0.05229 & 0.18378 \\ 
		All\_Sum\_FAMean & 213.26306 & 189.73499 & 0.00510 & $*$ \\ 
		All\_Sum\_FiberLengthMean & 16550.10867 & 14481.75686 & 0.00895 & $*$ \\ 
		All\_Sum\_FiberN & 10821.80000 & 10557.92857 & 0.15757 \\ 
		All\_Sum\_FiberNDivLength & 460.41013 & 467.08372 & 0.56140 \\ 
		All\_Sum\_Unweighted & 552.93333 & 530.78571 & 0.02924 \\ 
		Left\_AdjLMaxDivD\_FAMean & 1.33695 & 1.37581 & 0.01968 \\ 
		Left\_AdjLMaxDivD\_FiberLengthMean & 1.42381 & 1.39134 & 0.22194 \\ 
		Left\_AdjLMaxDivD\_FiberN & 1.99137 & 2.04882 & 0.35745 \\ 
		Left\_AdjLMaxDivD\_FiberNDivLength & 1.76567 & 1.79038 & 0.55969 \\ 
		Left\_AdjLMaxDivD\_Unweighted & 1.24266 & 1.24448 & 0.85912 \\ 
		Left\_HoffmanBound\_FAMean & 4.57409 & 4.41467 & 0.12060 \\ 
		Left\_HoffmanBound\_FiberLengthMean & 3.21652 & 3.27301 & 0.53075 \\ 
		Left\_HoffmanBound\_FiberN & 2.73899 & 2.63259 & 0.12089 \\ 
		Left\_HoffmanBound\_FiberNDivLength & 2.62682 & 2.66592 & 0.65419 \\ 
		Left\_HoffmanBound\_Unweighted & 4.68245 & 4.52121 & 0.04470 \\ 
		Left\_LogSpanningForestN\_FAMean & 52.24226 & 46.42507 & 0.00990 & $*$ \\ 
		Left\_LogSpanningForestN\_FiberLengthMean & 227.92206 & 223.08226 & 0.06568 \\ 
		Left\_LogSpanningForestN\_FiberN & 198.01945 & 195.98984 & 0.14698 \\ 
		Left\_LogSpanningForestN\_FiberNDivLength & 72.92756 & 73.02259 & 0.94641 \\ 
		Left\_LogSpanningForestN\_Unweighted & 94.40670 & 92.55840 & 0.12817 \\ 
		Left\_MaxFracMatching\_FAMean & 9.21339 & 8.44291 & 0.00192 & $*$ \\ 
		Left\_MaxFracMatching\_FiberLengthMean & 1063.15137 & 957.78591 & 0.04989 \\ 
		Left\_MaxFracMatching\_FiberN & 1143.30000 & 1167.75000 & 0.55643 \\ 
		Left\_MaxFracMatching\_FiberNDivLength & 54.05855 & 54.06370 & 0.99810 \\ 
		Left\_MaxFracMatching\_Unweighted & 20.73333 & 20.64286 & 0.44362 \\ 
		Left\_MaxMatching\_FAMean & 9.14568 & 8.41959 & 0.00408 & $*$ \\ 
		Left\_MaxMatching\_FiberLengthMean & 1069.71962 & 956.71517 & 0.04596 \\ 
		Left\_MaxMatching\_FiberN & 1132.06667 & 1173.28571 & 0.30854 \\ 
		Left\_MaxMatching\_FiberNDivLength & 53.32822 & 53.95062 & 0.76065 \\ 
		Left\_MaxMatching\_Unweighted & 20.46667 & 20.50000 & 0.86371 \\ 
		Left\_MinCutBalDivSum\_FAMean & 0.23163 & 0.21678 & 0.26822 \\ 
		Left\_MinCutBalDivSum\_FiberLengthMean & 0.21504 & 0.20569 & 0.54701 \\ 
		Left\_MinCutBalDivSum\_FiberN & 0.12105 & 0.11721 & 0.61607 \\ 
		Left\_MinCutBalDivSum\_FiberNDivLength & 0.12600 & 0.11970 & 0.34886 \\ 
		Left\_MinCutBalDivSum\_Unweighted & 0.23212 & 0.21513 & 0.15751 \\ 
		Left\_MinSpanningForest\_FAMean & 9.73295 & 8.62564 & 0.00644 & $*$ \\ 
		Left\_MinSpanningForest\_FiberLengthMean & 554.92086 & 550.46353 & 0.59531 \\ 
		Left\_MinSpanningForest\_FiberN & 51.20000 & 55.42857 & 0.14696 \\ 
		Left\_MinSpanningForest\_FiberNDivLength & 1.77670 & 2.12087 & 0.13439 \\ 
		Left\_MinVertexCoverBinary\_Unweighted & 30.13333 & 29.57143 & 0.20838 \\ 
		Left\_MinVertexCover\_FAMean & 9.21339 & 8.44291 & 0.00192 & $*$ \\ 
		Left\_MinVertexCover\_FiberLengthMean & 1063.15137 & 957.78591 & 0.04989 \\ 
		Left\_MinVertexCover\_FiberN & 1143.30000 & 1167.75000 & 0.55643 \\ 
		Left\_MinVertexCover\_FiberNDivLength & 54.05855 & 54.06370 & 0.99810 \\ 
		Left\_MinVertexCover\_Unweighted & 20.73333 & 20.64286 & 0.44362 \\ 
		Left\_PGEigengap\_FAMean & 0.30932 & 0.28263 & 0.18222 \\ 
		Left\_PGEigengap\_FiberLengthMean & 0.30530 & 0.27410 & 0.21739 \\ 
		Left\_PGEigengap\_FiberN & 0.15447 & 0.14019 & 0.17287 \\ 
		Left\_PGEigengap\_FiberNDivLength & 0.13316 & 0.12753 & 0.39949 \\ 
		Left\_PGEigengap\_Unweighted & 0.28522 & 0.25775 & 0.12577 \\ 
		Left\_Sum\_FAMean & 103.09873 & 91.35247 & 0.01578 \\ 
		Left\_Sum\_FiberLengthMean & 8389.35457 & 7367.36299 & 0.02104 \\ 
		Left\_Sum\_FiberN & 5377.13333 & 5240.85714 & 0.24514 \\ 
		Left\_Sum\_FiberNDivLength & 227.73780 & 231.42849 & 0.57511 \\ 
		Left\_Sum\_Unweighted & 274.93333 & 263.64286 & 0.07612 \\ 
		Right\_AdjLMaxDivD\_FAMean & 1.32925 & 1.35005 & 0.11677 \\ 
		Right\_AdjLMaxDivD\_FiberLengthMean & 1.42649 & 1.39061 & 0.21015 \\ 
		Right\_AdjLMaxDivD\_FiberN & 2.02880 & 2.12861 & 0.13979 \\ 
		Right\_AdjLMaxDivD\_FiberNDivLength & 1.77135 & 1.81936 & 0.16523 \\ 
		Right\_AdjLMaxDivD\_Unweighted & 1.24392 & 1.24183 & 0.83764 \\ 
		Right\_HoffmanBound\_FAMean & 4.33520 & 4.23355 & 0.24893 \\ 
		Right\_HoffmanBound\_FiberLengthMean & 3.36481 & 3.39929 & 0.72637 \\ 
		Right\_HoffmanBound\_FiberN & 2.64897 & 2.59370 & 0.43090 \\ 
		Right\_HoffmanBound\_FiberNDivLength & 2.51960 & 2.58206 & 0.48071 \\ 
		Right\_HoffmanBound\_Unweighted & 4.53947 & 4.48295 & 0.42576 \\ 
		Right\_LogSpanningForestN\_FAMean & 50.74546 & 45.62445 & 0.00422 & $*$ \\ 
		Right\_LogSpanningForestN\_FiberLengthMean & 218.55743 & 213.89157 & 0.03654 \\ 
		Right\_LogSpanningForestN\_FiberN & 189.81521 & 187.54735 & 0.07859 \\ 
		Right\_LogSpanningForestN\_FiberNDivLength & 68.78101 & 67.89394 & 0.44590 \\ 
		Right\_LogSpanningForestN\_Unweighted & 89.90052 & 87.96104 & 0.03973 \\ 
		Right\_MaxFracMatching\_FAMean & 9.16146 & 8.66941 & 0.03172 \\ 
		Right\_MaxFracMatching\_FiberLengthMean & 957.31955 & 845.29704 & 0.02148 \\ 
		Right\_MaxFracMatching\_FiberN & 1121.66667 & 1175.42857 & 0.17210 \\ 
		Right\_MaxFracMatching\_FiberNDivLength & 52.40863 & 54.91359 & 0.24701 \\ 
		Right\_MaxFracMatching\_Unweighted & 20.16667 & 20.21429 & 0.68799 \\ 
		Right\_MaxMatching\_FAMean & 9.11447 & 8.66452 & 0.05138 \\ 
		Right\_MaxMatching\_FiberLengthMean & 953.80876 & 844.31972 & 0.02578 \\ 
		Right\_MaxMatching\_FiberN & 1120.06667 & 1173.28571 & 0.18539 \\ 
		Right\_MaxMatching\_FiberNDivLength & 52.12333 & 55.05770 & 0.17603 \\ 
		Right\_MaxMatching\_Unweighted & 19.80000 & 20.00000 & 0.08221 \\ 
		Right\_MinCutBalDivSum\_FAMean & 0.23842 & 0.21535 & 0.01075 \\ 
		Right\_MinCutBalDivSum\_FiberLengthMean & 0.22416 & 0.20040 & 0.02688 \\ 
		Right\_MinCutBalDivSum\_FiberN & 0.13019 & 0.11780 & 0.02021 \\ 
		Right\_MinCutBalDivSum\_FiberNDivLength & 0.12662 & 0.12085 & 0.30422 \\ 
		Right\_MinCutBalDivSum\_Unweighted & 0.23090 & 0.20775 & 0.00292 & $*$ \\ 
		Right\_MinSpanningForest\_FAMean & 10.16814 & 9.35471 & 0.04285 \\ 
		Right\_MinSpanningForest\_FiberLengthMean & 543.52392 & 531.64092 & 0.20128 \\ 
		Right\_MinSpanningForest\_FiberN & 50.46667 & 52.07143 & 0.45408 \\ 
		Right\_MinSpanningForest\_FiberNDivLength & 1.82428 & 2.03587 & 0.12688 \\ 
		Right\_MinVertexCoverBinary\_Unweighted & 28.86667 & 28.78571 & 0.85585 \\ 
		Right\_MinVertexCover\_FAMean & 9.16146 & 8.66941 & 0.03172 \\ 
		Right\_MinVertexCover\_FiberLengthMean & 957.31955 & 845.29704 & 0.02148 \\ 
		Right\_MinVertexCover\_FiberN & 1121.66667 & 1175.42857 & 0.17210 \\ 
		Right\_MinVertexCover\_FiberNDivLength & 52.40863 & 54.91359 & 0.24701 \\ 
		Right\_MinVertexCover\_Unweighted & 20.16667 & 20.21429 & 0.68799 \\ 
		Right\_PGEigengap\_FAMean & 0.30913 & 0.26010 & 0.00059 & $*$ \\ 
		Right\_PGEigengap\_FiberLengthMean & 0.31935 & 0.24662 & 0.00031 & $*$ \\ 
		Right\_PGEigengap\_FiberN & 0.16550 & 0.13996 & 0.00766 & $*$ \\ 
		Right\_PGEigengap\_FiberNDivLength & 0.14237 & 0.12923 & 0.04494 \\ 
		Right\_PGEigengap\_Unweighted & 0.28041 & 0.23217 & 0.00007 & $*$ \\ 
		Right\_Sum\_FAMean & 101.47390 & 89.64221 & 0.00240 & $*$ \\ 
		Right\_Sum\_FiberLengthMean & 7658.04606 & 6588.59066 & 0.00616 & $*$ \\ 
		Right\_Sum\_FiberN & 5138.73333 & 5017.57143 & 0.29217 \\ 
		Right\_Sum\_FiberNDivLength & 218.26533 & 220.84907 & 0.65453 \\ 
		Right\_Sum\_Unweighted & 257.13333 & 244.42857 & 0.01112 \\ 
	\end{longtable}
	
}

\small{
	\subsection{129 nodes, round 1}
	\label{Table_Round1_60}
	\begin{longtable}{l | cccc}
		Property & Female & Male & p-value &  \\ 
		All\_AdjLMaxDivD\_FAMean & 1.41264 & 1.43719 & 0.17554 \\ 
		All\_AdjLMaxDivD\_FiberLengthMean & 1.51388 & 1.48806 & 0.38135 \\ 
		All\_AdjLMaxDivD\_FiberN & 2.16657 & 2.28753 & 0.15888 \\ 
		All\_AdjLMaxDivD\_FiberNDivLength & 2.03504 & 2.11052 & 0.32365 \\ 
		All\_AdjLMaxDivD\_Unweighted & 1.29906 & 1.29108 & 0.47042 \\ 
		All\_HoffmanBound\_FAMean & 4.34264 & 4.26186 & 0.14326 \\ 
		All\_HoffmanBound\_FiberLengthMean & 3.23922 & 3.23931 & 0.99908 \\ 
		All\_HoffmanBound\_FiberN & 2.50923 & 2.48179 & 0.69252 \\ 
		All\_HoffmanBound\_FiberNDivLength & 2.40134 & 2.40527 & 0.95531 \\ 
		All\_HoffmanBound\_Unweighted & 4.57882 & 4.48873 & 0.13289 \\ 
		All\_LeftRatio\_FAMean & 0.99644 & 0.98958 & 0.69917 \\ 
		All\_LeftRatio\_FiberLengthMean & 1.03985 & 1.02933 & 0.57321 \\ 
		All\_LeftRatio\_FiberN & 0.98707 & 0.98873 & 0.89300 \\ 
		All\_LeftRatio\_FiberNDivLength & 0.98306 & 0.98686 & 0.72266 \\ 
		All\_LeftRatio\_Unweighted & 1.01804 & 1.01148 & 0.52537 \\ 
		All\_LogSpanningForestN\_FAMean & 189.65029 & 175.14710 & 0.00714 & $*$ \\ 
		All\_LogSpanningForestN\_FiberLengthMean & 736.38352 & 723.03913 & 0.03944 \\ 
		All\_LogSpanningForestN\_FiberN & 596.77500 & 590.36663 & 0.02532 \\ 
		All\_LogSpanningForestN\_FiberNDivLength & 208.78477 & 208.06750 & 0.79579 \\ 
		All\_LogSpanningForestN\_Unweighted & 319.35263 & 315.28948 & 0.14715 \\ 
		All\_MaxFracMatching\_FAMean & 47.79299 & 47.73239 & 0.99309 \\ 
		All\_MaxFracMatching\_FiberLengthMean & 3255.53791 & 2885.46483 & 0.01031 \\ 
		All\_MaxFracMatching\_FiberN & 2450.83333 & 2354.17857 & 0.10402 \\ 
		All\_MaxFracMatching\_FiberNDivLength & 131.20093 & 128.48485 & 0.60374 \\ 
		All\_MaxFracMatching\_Unweighted & 63.90000 & 63.82143 & 0.62381 \\ 
		All\_MaxMatching\_FAMean & 47.61671 & 47.42748 & 0.97825 \\ 
		All\_MaxMatching\_FiberLengthMean & 3250.31191 & 2881.03276 & 0.01010 \\ 
		All\_MaxMatching\_FiberN & 2442.06667 & 2349.00000 & 0.11996 \\ 
		All\_MaxMatching\_FiberNDivLength & 130.78627 & 128.19811 & 0.61957 \\ 
		All\_MaxMatching\_Unweighted & 63.60000 & 63.50000 & 0.60403 \\ 
		All\_MinCutBalDivSum\_FAMean & 0.04107 & 0.05867 & 0.03314 \\ 
		All\_MinCutBalDivSum\_FiberLengthMean & 0.01648 & 0.02001 & 0.23049 \\ 
		All\_MinCutBalDivSum\_FiberN & 0.02515 & 0.02466 & 0.87507 \\ 
		All\_MinCutBalDivSum\_FiberNDivLength & 0.04344 & 0.05116 & 0.26759 \\ 
		All\_MinCutBalDivSum\_Unweighted & 0.02012 & 0.02359 & 0.09504 \\ 
		All\_MinSpanningForest\_FAMean & 29.80087 & 27.32132 & 0.00560 & $*$ \\ 
		All\_MinSpanningForest\_FiberLengthMean & 1660.12748 & 1619.59859 & 0.00419 & $*$ \\ 
		All\_MinSpanningForest\_FiberN & 140.73333 & 139.35714 & 0.42330 \\ 
		All\_MinSpanningForest\_FiberNDivLength & 4.46488 & 4.57888 & 0.56445 \\ 
		All\_MinVertexCoverBinary\_Unweighted & 95.60000 & 95.50000 & 0.86573 \\ 
		All\_MinVertexCover\_FAMean & 29.28164 & 27.54695 & 0.00867 & $*$ \\ 
		All\_MinVertexCover\_FiberLengthMean & 3254.28136 & 2885.14192 & 0.01053 \\ 
		All\_MinVertexCover\_FiberN & 2450.83333 & 2354.17857 & 0.10402 \\ 
		All\_MinVertexCover\_FiberNDivLength & 122.18259 & 120.33442 & 0.64280 \\ 
		All\_MinVertexCover\_Unweighted & 63.90000 & 63.82143 & 0.62381 \\ 
		All\_PGEigengap\_FAMean & 0.02962 & 0.03269 & 0.30221 \\ 
		All\_PGEigengap\_FiberLengthMean & 0.02299 & 0.02684 & 0.28803 \\ 
		All\_PGEigengap\_FiberN & 0.02534 & 0.02418 & 0.58917 \\ 
		All\_PGEigengap\_FiberNDivLength & 0.02574 & 0.02491 & 0.67919 \\ 
		All\_PGEigengap\_Unweighted & 0.02738 & 0.03019 & 0.25874 \\ 
		All\_Sum\_FAMean & 383.21280 & 343.50552 & 0.00566 & $*$ \\ 
		All\_Sum\_FiberLengthMean & 29954.75727 & 26359.29258 & 0.01648 \\ 
		All\_Sum\_FiberN & 12055.20000 & 11753.07143 & 0.08683 \\ 
		All\_Sum\_FiberNDivLength & 539.10743 & 543.51648 & 0.72060 \\ 
		All\_Sum\_Unweighted & 996.80000 & 960.07143 & 0.05695 \\ 
		Left\_AdjLMaxDivD\_FAMean & 1.38282 & 1.41735 & 0.01972 \\ 
		Left\_AdjLMaxDivD\_FiberLengthMean & 1.44089 & 1.42883 & 0.57130 \\ 
		Left\_AdjLMaxDivD\_FiberN & 1.88927 & 2.02446 & 0.08956 \\ 
		Left\_AdjLMaxDivD\_FiberNDivLength & 1.77179 & 1.86037 & 0.15990 \\ 
		Left\_AdjLMaxDivD\_Unweighted & 1.26569 & 1.26226 & 0.67723 \\ 
		Left\_HoffmanBound\_FAMean & 4.55896 & 4.42698 & 0.04890 \\ 
		Left\_HoffmanBound\_FiberLengthMean & 3.28256 & 3.25849 & 0.74494 \\ 
		Left\_HoffmanBound\_FiberN & 2.77133 & 2.69944 & 0.31545 \\ 
		Left\_HoffmanBound\_FiberNDivLength & 2.65783 & 2.64829 & 0.90792 \\ 
		Left\_HoffmanBound\_Unweighted & 4.74063 & 4.59434 & 0.01729 \\ 
		Left\_LogSpanningForestN\_FAMean & 94.16378 & 85.56169 & 0.00401 & $*$ \\ 
		Left\_LogSpanningForestN\_FiberLengthMean & 370.60101 & 362.50884 & 0.02008 \\ 
		Left\_LogSpanningForestN\_FiberN & 299.54862 & 296.04072 & 0.04778 \\ 
		Left\_LogSpanningForestN\_FiberNDivLength & 104.45062 & 104.20143 & 0.88960 \\ 
		Left\_LogSpanningForestN\_Unweighted & 160.53642 & 157.70585 & 0.06838 \\ 
		Left\_MaxFracMatching\_FAMean & 23.96060 & 23.75461 & 0.95447 \\ 
		Left\_MaxFracMatching\_FiberLengthMean & 1684.63258 & 1492.07886 & 0.01201 \\ 
		Left\_MaxFracMatching\_FiberN & 1158.83333 & 1143.71429 & 0.70301 \\ 
		Left\_MaxFracMatching\_FiberNDivLength & 62.85453 & 63.36024 & 0.86300 \\ 
		Left\_MaxFracMatching\_Unweighted & 32.23333 & 32.14286 & 0.44362 \\ 
		Left\_MaxMatching\_FAMean & 23.78976 & 23.59561 & 0.95648 \\ 
		Left\_MaxMatching\_FiberLengthMean & 1681.84910 & 1490.02843 & 0.01201 \\ 
		Left\_MaxMatching\_FiberN & 1155.06667 & 1140.78571 & 0.71945 \\ 
		Left\_MaxMatching\_FiberNDivLength & 62.56412 & 63.10921 & 0.85180 \\ 
		Left\_MaxMatching\_Unweighted & 32.00000 & 31.85714 & 0.13873 \\ 
		Left\_MinCutBalDivSum\_FAMean & 0.39142 & 0.42795 & 0.51193 \\ 
		Left\_MinCutBalDivSum\_FiberLengthMean & 0.18373 & 0.17722 & 0.63821 \\ 
		Left\_MinCutBalDivSum\_FiberN & 0.11504 & 0.10370 & 0.17204 \\ 
		Left\_MinCutBalDivSum\_FiberNDivLength & 0.29372 & 0.28518 & 0.87111 \\ 
		Left\_MinCutBalDivSum\_Unweighted & 0.18528 & 0.17527 & 0.31089 \\ 
		Left\_MinSpanningForest\_FAMean & 14.31647 & 13.21554 & 0.02337 \\ 
		Left\_MinSpanningForest\_FiberLengthMean & 832.18125 & 816.19078 & 0.08567 \\ 
		Left\_MinSpanningForest\_FiberN & 69.93333 & 71.50000 & 0.53489 \\ 
		Left\_MinSpanningForest\_FiberNDivLength & 2.19041 & 2.40779 & 0.30460 \\ 
		Left\_MinVertexCoverBinary\_Unweighted & 48.60000 & 48.07143 & 0.26704 \\ 
		Left\_MinVertexCover\_FAMean & 14.58465 & 13.49574 & 0.00217 & $*$ \\ 
		Left\_MinVertexCover\_FiberLengthMean & 1683.88816 & 1491.95077 & 0.01217 \\ 
		Left\_MinVertexCover\_FiberN & 1158.83333 & 1143.71429 & 0.70301 \\ 
		Left\_MinVertexCover\_FiberNDivLength & 58.18846 & 59.30983 & 0.63976 \\ 
		Left\_MinVertexCover\_Unweighted & 32.23333 & 32.14286 & 0.44362 \\ 
		Left\_PGEigengap\_FAMean & 0.20897 & 0.19062 & 0.26005 \\ 
		Left\_PGEigengap\_FiberLengthMean & 0.21073 & 0.19264 & 0.37328 \\ 
		Left\_PGEigengap\_FiberN & 0.11048 & 0.09812 & 0.17273 \\ 
		Left\_PGEigengap\_FiberNDivLength & 0.08678 & 0.08249 & 0.37534 \\ 
		Left\_PGEigengap\_Unweighted & 0.18680 & 0.16841 & 0.20490 \\ 
		Left\_Sum\_FAMean & 190.73838 & 169.91048 & 0.00640 & $*$ \\ 
		Left\_Sum\_FiberLengthMean & 15551.42767 & 13553.58981 & 0.01063 \\ 
		Left\_Sum\_FiberN & 5953.20000 & 5808.28571 & 0.23195 \\ 
		Left\_Sum\_FiberNDivLength & 265.28431 & 268.10642 & 0.70300 \\ 
		Left\_Sum\_Unweighted & 507.13333 & 485.50000 & 0.03150 \\ 
		Right\_AdjLMaxDivD\_FAMean & 1.36927 & 1.36644 & 0.86545 \\ 
		Right\_AdjLMaxDivD\_FiberLengthMean & 1.46663 & 1.42300 & 0.22845 \\ 
		Right\_AdjLMaxDivD\_FiberN & 2.11492 & 2.26216 & 0.13799 \\ 
		Right\_AdjLMaxDivD\_FiberNDivLength & 1.81282 & 1.88803 & 0.20555 \\ 
		Right\_AdjLMaxDivD\_Unweighted & 1.26402 & 1.24367 & 0.05722 \\ 
		Right\_HoffmanBound\_FAMean & 4.31976 & 4.24036 & 0.35747 \\ 
		Right\_HoffmanBound\_FiberLengthMean & 3.31782 & 3.39672 & 0.40216 \\ 
		Right\_HoffmanBound\_FiberN & 2.63936 & 2.53312 & 0.14213 \\ 
		Right\_HoffmanBound\_FiberNDivLength & 2.60907 & 2.64209 & 0.57976 \\ 
		Right\_HoffmanBound\_Unweighted & 4.54479 & 4.47165 & 0.33936 \\ 
		Right\_LogSpanningForestN\_FAMean & 90.78121 & 84.79885 & 0.05405 \\ 
		Right\_LogSpanningForestN\_FiberLengthMean & 357.29779 & 351.79798 & 0.14316 \\ 
		Right\_LogSpanningForestN\_FiberN & 289.78041 & 286.90175 & 0.12791 \\ 
		Right\_LogSpanningForestN\_FiberNDivLength & 99.92977 & 99.41627 & 0.73502 \\ 
		Right\_LogSpanningForestN\_Unweighted & 153.35116 & 151.79460 & 0.36897 \\ 
		Right\_MaxFracMatching\_FAMean & 23.64340 & 23.92987 & 0.93393 \\ 
		Right\_MaxFracMatching\_FiberLengthMean & 1550.30409 & 1379.09472 & 0.02294 \\ 
		Right\_MaxFracMatching\_FiberN & 1211.23333 & 1160.21429 & 0.17444 \\ 
		Right\_MaxFracMatching\_FiberNDivLength & 66.22967 & 64.18666 & 0.48841 \\ 
		Right\_MaxFracMatching\_Unweighted & 31.66667 & 31.67857 & 0.92581 \\ 
		Right\_MaxMatching\_FAMean & 23.56597 & 23.74101 & 0.95917 \\ 
		Right\_MaxMatching\_FiberLengthMean & 1547.55246 & 1376.37203 & 0.02307 \\ 
		Right\_MaxMatching\_FiberN & 1206.66667 & 1157.00000 & 0.18644 \\ 
		Right\_MaxMatching\_FiberNDivLength & 66.04844 & 63.96700 & 0.47644 \\ 
		Right\_MaxMatching\_Unweighted & 31.53333 & 31.35714 & 0.35824 \\ 
		Right\_MinCutBalDivSum\_FAMean & 0.37749 & 0.40030 & 0.65833 \\ 
		Right\_MinCutBalDivSum\_FiberLengthMean & 0.18310 & 0.16294 & 0.08674 \\ 
		Right\_MinCutBalDivSum\_FiberN & 0.10626 & 0.09805 & 0.06373 \\ 
		Right\_MinCutBalDivSum\_FiberNDivLength & 0.27529 & 0.26502 & 0.82417 \\ 
		Right\_MinCutBalDivSum\_Unweighted & 0.18178 & 0.16539 & 0.05282 \\ 
		Right\_MinSpanningForest\_FAMean & 15.54990 & 14.25185 & 0.02184 \\ 
		Right\_MinSpanningForest\_FiberLengthMean & 822.88285 & 798.47914 & 0.01294 \\ 
		Right\_MinSpanningForest\_FiberN & 71.06667 & 69.64286 & 0.33198 \\ 
		Right\_MinSpanningForest\_FiberNDivLength & 2.33231 & 2.35222 & 0.88466 \\ 
		Right\_MinVertexCoverBinary\_Unweighted & 47.00000 & 47.07143 & 0.87068 \\ 
		Right\_MinVertexCover\_FAMean & 14.49525 & 13.89677 & 0.09572 \\ 
		Right\_MinVertexCover\_FiberLengthMean & 1550.06787 & 1378.66029 & 0.02307 \\ 
		Right\_MinVertexCover\_FiberN & 1211.23333 & 1160.21429 & 0.17444 \\ 
		Right\_MinVertexCover\_FiberNDivLength & 61.99819 & 60.10934 & 0.42599 \\ 
		Right\_MinVertexCover\_Unweighted & 31.66667 & 31.67857 & 0.92581 \\ 
		Right\_PGEigengap\_FAMean & 0.21510 & 0.18486 & 0.01701 \\ 
		Right\_PGEigengap\_FiberLengthMean & 0.21993 & 0.17854 & 0.01790 \\ 
		Right\_PGEigengap\_FiberN & 0.11815 & 0.09973 & 0.00710 & $*$ \\ 
		Right\_PGEigengap\_FiberNDivLength & 0.09679 & 0.08770 & 0.03556 \\ 
		Right\_PGEigengap\_Unweighted & 0.19093 & 0.16211 & 0.01242 \\ 
		Right\_Sum\_FAMean & 183.61218 & 164.74653 & 0.01787 \\ 
		Right\_Sum\_FiberLengthMean & 13876.53350 & 12267.36700 & 0.03808 \\ 
		Right\_Sum\_FiberN & 5795.86667 & 5648.07143 & 0.17406 \\ 
		Right\_Sum\_FiberNDivLength & 259.38730 & 260.72185 & 0.82866 \\ 
		Right\_Sum\_Unweighted & 468.46667 & 451.42857 & 0.13292 \\ 
	\end{longtable}
	
}

\small{
	\subsection{234 nodes, round 1}
	\label{Table_Round1_125}
	\begin{longtable}{l | cccc}
		Property & Female & Male & p-value &  \\ 
		All\_AdjLMaxDivD\_FAMean & 1.59803 & 1.64024 & 0.08803 \\ 
		All\_AdjLMaxDivD\_FiberLengthMean & 1.73972 & 1.72166 & 0.69642 \\ 
		All\_AdjLMaxDivD\_FiberN & 2.98297 & 3.17562 & 0.18871 \\ 
		All\_AdjLMaxDivD\_FiberNDivLength & 2.90553 & 3.05789 & 0.32675 \\ 
		All\_AdjLMaxDivD\_Unweighted & 1.43488 & 1.44187 & 0.69379 \\ 
		All\_HoffmanBound\_FAMean & 4.05843 & 4.02343 & 0.43528 \\ 
		All\_HoffmanBound\_FiberLengthMean & 3.10090 & 3.12089 & 0.78071 \\ 
		All\_HoffmanBound\_FiberN & 2.37479 & 2.35909 & 0.81574 \\ 
		All\_HoffmanBound\_FiberNDivLength & 2.32613 & 2.31582 & 0.89211 \\ 
		All\_HoffmanBound\_Unweighted & 4.24417 & 4.19579 & 0.33435 \\ 
		All\_LeftRatio\_FAMean & 0.99382 & 0.98754 & 0.70349 \\ 
		All\_LeftRatio\_FiberLengthMean & 1.03212 & 1.01751 & 0.39210 \\ 
		All\_LeftRatio\_FiberN & 0.99604 & 0.99866 & 0.82533 \\ 
		All\_LeftRatio\_FiberNDivLength & 0.99635 & 1.00000 & 0.73087 \\ 
		All\_LeftRatio\_Unweighted & 1.01658 & 1.00968 & 0.49891 \\ 
		All\_LogSpanningForestN\_FAMean & 323.14013 & 299.07957 & 0.00428 & $*$ \\ 
		All\_LogSpanningForestN\_FiberLengthMean & 1313.09259 & 1289.79813 & 0.03407 \\ 
		All\_LogSpanningForestN\_FiberN & 954.63726 & 942.88185 & 0.04603 \\ 
		All\_LogSpanningForestN\_FiberNDivLength & 260.95250 & 258.64163 & 0.65016 \\ 
		All\_LogSpanningForestN\_Unweighted & 569.99349 & 562.00182 & 0.11783 \\ 
		All\_MaxFracMatching\_FAMean & 77.47689 & 86.69052 & 0.47644 \\ 
		All\_MaxFracMatching\_FiberLengthMean & 5264.48143 & 4669.21300 & 0.00595 & $*$ \\ 
		All\_MaxFracMatching\_FiberN & 2423.03333 & 2346.85714 & 0.02910 \\ 
		All\_MaxFracMatching\_FiberNDivLength & 146.84213 & 152.18921 & 0.50458 \\ 
		All\_MaxFracMatching\_Unweighted & 116.30000 & 116.10714 & 0.31677 \\ 
		All\_MaxMatching\_FAMean & 77.32548 & 86.50399 & 0.47709 \\ 
		All\_MaxMatching\_FiberLengthMean & 5267.90205 & 4664.13177 & 0.00501 & $*$ \\ 
		All\_MaxMatching\_FiberN & 2415.06667 & 2343.78571 & 0.03922 \\ 
		All\_MaxMatching\_FiberNDivLength & 146.14385 & 151.83590 & 0.47885 \\ 
		All\_MaxMatching\_Unweighted & 116.06667 & 115.85714 & 0.37721 \\ 
		All\_MinCutBalDivSum\_FAMean & 0.01514 & 0.02013 & 0.31079 \\ 
		All\_MinCutBalDivSum\_FiberLengthMean & 0.00958 & 0.01180 & 0.19575 \\ 
		All\_MinCutBalDivSum\_FiberN & 0.02351 & 0.02307 & 0.88283 \\ 
		All\_MinCutBalDivSum\_FiberNDivLength & 0.02792 & 0.03320 & 0.38016 \\ 
		All\_MinCutBalDivSum\_Unweighted & 0.01155 & 0.01338 & 0.14262 \\ 
		All\_MinSpanningForest\_FAMean & 50.15919 & 47.66875 & 0.03437 \\ 
		All\_MinSpanningForest\_FiberLengthMean & 2816.89991 & 2786.31135 & 0.05718 \\ 
		All\_MinSpanningForest\_FiberN & 246.66667 & 244.21429 & 0.32290 \\ 
		All\_MinSpanningForest\_FiberNDivLength & 8.09907 & 8.42327 & 0.35978 \\ 
		All\_MinVertexCoverBinary\_Unweighted & 166.46667 & 164.50000 & 0.14702 \\ 
		All\_MinVertexCover\_FAMean & 51.34924 & 48.31155 & 0.00298 & $*$ \\ 
		All\_MinVertexCover\_FiberLengthMean & 5263.01339 & 4680.07605 & 0.00624 & $*$ \\ 
		All\_MinVertexCover\_FiberN & 2429.30000 & 2347.64286 & 0.02321 \\ 
		All\_MinVertexCover\_FiberNDivLength & 129.96605 & 128.35597 & 0.58930 \\ 
		All\_MinVertexCover\_Unweighted & 116.26667 & 116.10714 & 0.40734 \\ 
		All\_PGEigengap\_FAMean & 0.01741 & 0.01917 & 0.32097 \\ 
		All\_PGEigengap\_FiberLengthMean & 0.01363 & 0.01606 & 0.25767 \\ 
		All\_PGEigengap\_FiberN & 0.02278 & 0.02175 & 0.59722 \\ 
		All\_PGEigengap\_FiberNDivLength & 0.02185 & 0.02112 & 0.67831 \\ 
		All\_PGEigengap\_Unweighted & 0.01558 & 0.01717 & 0.28447 \\ 
		All\_Sum\_FAMean & 663.03525 & 600.56092 & 0.00295 & $*$ \\ 
		All\_Sum\_FiberLengthMean & 50301.63361 & 44901.15736 & 0.01494 \\ 
		All\_Sum\_FiberN & 13010.73333 & 12717.64286 & 0.07686 \\ 
		All\_Sum\_FiberNDivLength & 612.72529 & 617.69320 & 0.69765 \\ 
		All\_Sum\_Unweighted & 1779.53333 & 1720.35714 & 0.05751 \\ 
		Left\_AdjLMaxDivD\_FAMean & 1.58188 & 1.64049 & 0.00188 & $*$ \\ 
		Left\_AdjLMaxDivD\_FiberLengthMean & 1.66505 & 1.67734 & 0.65526 \\ 
		Left\_AdjLMaxDivD\_FiberN & 2.53544 & 2.73240 & 0.07849 \\ 
		Left\_AdjLMaxDivD\_FiberNDivLength & 2.41172 & 2.58587 & 0.10342 \\ 
		Left\_AdjLMaxDivD\_Unweighted & 1.41220 & 1.42766 & 0.19979 \\ 
		Left\_HoffmanBound\_FAMean & 4.18213 & 4.15436 & 0.69068 \\ 
		Left\_HoffmanBound\_FiberLengthMean & 3.16103 & 3.13165 & 0.69090 \\ 
		Left\_HoffmanBound\_FiberN & 2.61131 & 2.59028 & 0.69022 \\ 
		Left\_HoffmanBound\_FiberNDivLength & 2.55253 & 2.52149 & 0.68292 \\ 
		Left\_HoffmanBound\_Unweighted & 4.36085 & 4.27881 & 0.19873 \\ 
		Left\_LogSpanningForestN\_FAMean & 160.55086 & 145.36374 & 0.00043 & $*$ \\ 
		Left\_LogSpanningForestN\_FiberLengthMean & 666.51767 & 651.53600 & 0.00282 & $*$ \\ 
		Left\_LogSpanningForestN\_FiberN & 483.77617 & 475.96167 & 0.01381 \\ 
		Left\_LogSpanningForestN\_FiberNDivLength & 131.44914 & 129.06273 & 0.45998 \\ 
		Left\_LogSpanningForestN\_Unweighted & 288.07253 & 282.09224 & 0.01055 \\ 
		Left\_MaxFracMatching\_FAMean & 39.10063 & 43.81774 & 0.48113 \\ 
		Left\_MaxFracMatching\_FiberLengthMean & 2723.34243 & 2404.61807 & 0.00298 & $*$ \\ 
		Left\_MaxFracMatching\_FiberN & 1190.10000 & 1185.96429 & 0.87482 \\ 
		Left\_MaxFracMatching\_FiberNDivLength & 74.54560 & 77.43346 & 0.48187 \\ 
		Left\_MaxFracMatching\_Unweighted & 59.13333 & 59.07143 & 0.67471 \\ 
		Left\_MaxMatching\_FAMean & 38.96026 & 43.61191 & 0.48363 \\ 
		Left\_MaxMatching\_FiberLengthMean & 2733.15685 & 2401.62800 & 0.00186 & $*$ \\ 
		Left\_MaxMatching\_FiberN & 1183.86667 & 1183.50000 & 0.98840 \\ 
		Left\_MaxMatching\_FiberNDivLength & 73.98551 & 77.20181 & 0.43128 \\ 
		Left\_MaxMatching\_Unweighted & 58.93333 & 58.78571 & 0.26516 \\ 
		Left\_MinCutBalDivSum\_FAMean & 0.17682 & 0.18449 & 0.88746 \\ 
		Left\_MinCutBalDivSum\_FiberLengthMean & 0.13085 & 0.12088 & 0.40639 \\ 
		Left\_MinCutBalDivSum\_FiberN & 0.09471 & 0.08283 & 0.10418 \\ 
		Left\_MinCutBalDivSum\_FiberNDivLength & 0.17229 & 0.17870 & 0.92359 \\ 
		Left\_MinCutBalDivSum\_Unweighted & 0.13170 & 0.12493 & 0.40732 \\ 
		Left\_MinSpanningForest\_FAMean & 24.84043 & 23.74120 & 0.12963 \\ 
		Left\_MinSpanningForest\_FiberLengthMean & 1429.01473 & 1418.56487 & 0.23304 \\ 
		Left\_MinSpanningForest\_FiberN & 128.13333 & 128.28571 & 0.95846 \\ 
		Left\_MinSpanningForest\_FiberNDivLength & 4.16924 & 4.54566 & 0.17006 \\ 
		Left\_MinVertexCoverBinary\_Unweighted & 84.26667 & 82.64286 & 0.03625 \\ 
		Left\_MinVertexCover\_FAMean & 25.73738 & 23.87149 & 0.00062 & $*$ \\ 
		Left\_MinVertexCover\_FiberLengthMean & 2723.68339 & 2414.02175 & 0.00300 & $*$ \\ 
		Left\_MinVertexCover\_FiberN & 1191.86667 & 1183.46429 & 0.75554 \\ 
		Left\_MinVertexCover\_FiberNDivLength & 65.88611 & 66.10788 & 0.90947 \\ 
		Left\_MinVertexCover\_Unweighted & 59.13333 & 59.03571 & 0.52377 \\ 
		Left\_PGEigengap\_FAMean & 0.13049 & 0.12036 & 0.34190 \\ 
		Left\_PGEigengap\_FiberLengthMean & 0.13090 & 0.12097 & 0.44738 \\ 
		Left\_PGEigengap\_FiberN & 0.08739 & 0.07658 & 0.11542 \\ 
		Left\_PGEigengap\_FiberNDivLength & 0.06688 & 0.06230 & 0.21569 \\ 
		Left\_PGEigengap\_Unweighted & 0.11641 & 0.10433 & 0.20469 \\ 
		Left\_Sum\_FAMean & 329.24155 & 296.86391 & 0.00264 & $*$ \\ 
		Left\_Sum\_FiberLengthMean & 25886.20314 & 22888.19605 & 0.00802 & $*$ \\ 
		Left\_Sum\_FiberN & 6488.00000 & 6345.21429 & 0.21051 \\ 
		Left\_Sum\_FiberNDivLength & 305.60029 & 308.85699 & 0.67183 \\ 
		Left\_Sum\_Unweighted & 902.66667 & 869.14286 & 0.02715 \\ 
		Right\_AdjLMaxDivD\_FAMean & 1.53535 & 1.54017 & 0.86686 \\ 
		Right\_AdjLMaxDivD\_FiberLengthMean & 1.66926 & 1.62871 & 0.45133 \\ 
		Right\_AdjLMaxDivD\_FiberN & 2.65172 & 2.83822 & 0.20338 \\ 
		Right\_AdjLMaxDivD\_FiberNDivLength & 2.33381 & 2.42285 & 0.35255 \\ 
		Right\_AdjLMaxDivD\_Unweighted & 1.39665 & 1.37491 & 0.26268 \\ 
		Right\_HoffmanBound\_FAMean & 4.09368 & 4.05022 & 0.42998 \\ 
		Right\_HoffmanBound\_FiberLengthMean & 3.16159 & 3.26745 & 0.17945 \\ 
		Right\_HoffmanBound\_FiberN & 2.54589 & 2.48387 & 0.22658 \\ 
		Right\_HoffmanBound\_FiberNDivLength & 2.54987 & 2.56558 & 0.76972 \\ 
		Right\_HoffmanBound\_Unweighted & 4.27328 & 4.24321 & 0.55020 \\ 
		Right\_LogSpanningForestN\_FAMean & 158.11129 & 149.15728 & 0.09220 \\ 
		Right\_LogSpanningForestN\_FiberLengthMean & 638.28386 & 629.75250 & 0.21522 \\ 
		Right\_LogSpanningForestN\_FiberN & 463.35779 & 459.45680 & 0.33133 \\ 
		Right\_LogSpanningForestN\_FiberNDivLength & 125.05479 & 125.10653 & 0.98601 \\ 
		Right\_LogSpanningForestN\_Unweighted & 276.64016 & 274.35545 & 0.50947 \\ 
		Right\_MaxFracMatching\_FAMean & 38.29213 & 42.79551 & 0.47324 \\ 
		Right\_MaxFracMatching\_FiberLengthMean & 2525.80952 & 2252.33000 & 0.02297 \\ 
		Right\_MaxFracMatching\_FiberN & 1143.36667 & 1114.50000 & 0.29332 \\ 
		Right\_MaxFracMatching\_FiberNDivLength & 69.99586 & 73.89356 & 0.37323 \\ 
		Right\_MaxFracMatching\_Unweighted & 57.16667 & 57.03571 & 0.37184 \\ 
		Right\_MaxMatching\_FAMean & 38.08601 & 42.66424 & 0.46269 \\ 
		Right\_MaxMatching\_FiberLengthMean & 2518.15458 & 2250.19058 & 0.02492 \\ 
		Right\_MaxMatching\_FiberN & 1143.93333 & 1112.92857 & 0.25811 \\ 
		Right\_MaxMatching\_FiberNDivLength & 69.89470 & 73.68388 & 0.38458 \\ 
		Right\_MaxMatching\_Unweighted & 56.73333 & 56.78571 & 0.75261 \\ 
		Right\_MinCutBalDivSum\_FAMean & 0.16974 & 0.17039 & 0.98933 \\ 
		Right\_MinCutBalDivSum\_FiberLengthMean & 0.13208 & 0.11282 & 0.05874 \\ 
		Right\_MinCutBalDivSum\_FiberN & 0.09241 & 0.08311 & 0.05924 \\ 
		Right\_MinCutBalDivSum\_FiberNDivLength & 0.16032 & 0.18258 & 0.71333 \\ 
		Right\_MinCutBalDivSum\_Unweighted & 0.12577 & 0.11149 & 0.05167 \\ 
		Right\_MinSpanningForest\_FAMean & 25.39625 & 24.04691 & 0.07033 \\ 
		Right\_MinSpanningForest\_FiberLengthMean & 1380.70294 & 1360.83076 & 0.09218 \\ 
		Right\_MinSpanningForest\_FiberN & 118.93333 & 117.35714 & 0.22005 \\ 
		Right\_MinSpanningForest\_FiberNDivLength & 3.97662 & 4.03133 & 0.77653 \\ 
		Right\_MinVertexCoverBinary\_Unweighted & 81.73333 & 81.35714 & 0.64589 \\ 
		Right\_MinVertexCover\_FAMean & 25.47790 & 24.31020 & 0.04371 \\ 
		Right\_MinVertexCover\_FiberLengthMean & 2524.64124 & 2253.53347 & 0.02391 \\ 
		Right\_MinVertexCover\_FiberN & 1147.73333 & 1115.07143 & 0.24439 \\ 
		Right\_MinVertexCover\_FiberNDivLength & 62.01048 & 61.37838 & 0.76647 \\ 
		Right\_MinVertexCover\_Unweighted & 57.13333 & 57.07143 & 0.67471 \\ 
		Right\_PGEigengap\_FAMean & 0.13264 & 0.10599 & 0.00210 & $*$ \\ 
		Right\_PGEigengap\_FiberLengthMean & 0.14008 & 0.10493 & 0.00216 & $*$ \\ 
		Right\_PGEigengap\_FiberN & 0.09018 & 0.07181 & 0.00216 & $*$ \\ 
		Right\_PGEigengap\_FiberNDivLength & 0.07040 & 0.06006 & 0.00239 & $*$ \\ 
		Right\_PGEigengap\_Unweighted & 0.11555 & 0.09106 & 0.00138 & $*$ \\ 
		Right\_Sum\_FAMean & 325.09144 & 294.90648 & 0.01679 \\ 
		Right\_Sum\_FiberLengthMean & 23910.27439 & 21483.83489 & 0.04481 \\ 
		Right\_Sum\_FiberN & 6217.00000 & 6073.57143 & 0.22048 \\ 
		Right\_Sum\_FiberNDivLength & 292.73367 & 294.05788 & 0.84987 \\ 
		Right\_Sum\_Unweighted & 855.93333 & 828.42857 & 0.17741 \\ 
	\end{longtable}
	
}

\small{
	\subsection{463 nodes, round 1}
	\label{Table_Round1_250}
	\begin{longtable}{l | cccc}
		Property & Female & Male & p-value &  \\ 
		All\_AdjLMaxDivD\_FAMean & 2.15215 & 2.18281 & 0.42513 \\ 
		All\_AdjLMaxDivD\_FiberLengthMean & 2.36923 & 2.34378 & 0.69634 \\ 
		All\_AdjLMaxDivD\_FiberN & 5.04437 & 5.35726 & 0.28501 \\ 
		All\_AdjLMaxDivD\_FiberNDivLength & 4.86199 & 5.11531 & 0.39426 \\ 
		All\_AdjLMaxDivD\_Unweighted & 1.88052 & 1.87368 & 0.82914 \\ 
		All\_HoffmanBound\_FAMean & 3.65420 & 3.58216 & 0.11960 \\ 
		All\_HoffmanBound\_FiberLengthMean & 2.96234 & 2.96941 & 0.90399 \\ 
		All\_HoffmanBound\_FiberN & 2.28012 & 2.25731 & 0.67430 \\ 
		All\_HoffmanBound\_FiberNDivLength & 2.25237 & 2.24638 & 0.91985 \\ 
		All\_HoffmanBound\_Unweighted & 3.74999 & 3.70604 & 0.31877 \\ 
		All\_LeftRatio\_FAMean & 0.98244 & 0.97722 & 0.77315 \\ 
		All\_LeftRatio\_FiberLengthMean & 1.01576 & 1.00512 & 0.55466 \\ 
		All\_LeftRatio\_FiberN & 0.99557 & 0.99767 & 0.86144 \\ 
		All\_LeftRatio\_FiberNDivLength & 0.99550 & 0.99827 & 0.79576 \\ 
		All\_LeftRatio\_Unweighted & 1.00742 & 0.99995 & 0.52098 \\ 
		All\_LogSpanningForestN\_FAMean & 432.44074 & 391.29116 & 0.01203 \\ 
		All\_LogSpanningForestN\_FiberLengthMean & 2312.59245 & 2271.99890 & 0.12674 \\ 
		All\_LogSpanningForestN\_FiberN & 1455.41884 & 1430.04191 & 0.08913 \\ 
		All\_LogSpanningForestN\_FiberNDivLength & 151.41709 & 146.09748 & 0.60236 \\ 
		All\_LogSpanningForestN\_Unweighted & 934.57522 & 921.32293 & 0.33372 \\ 
		All\_MaxFracMatching\_FAMean & 97.96938 & 83.39334 & 0.12442 \\ 
		All\_MaxFracMatching\_FiberLengthMean & 8045.71403 & 7329.01371 & 0.02703 \\ 
		All\_MaxFracMatching\_FiberN & 2447.73333 & 2363.89286 & 0.01650 \\ 
		All\_MaxFracMatching\_FiberNDivLength & 139.63506 & 129.89153 & 0.27253 \\ 
		All\_MaxFracMatching\_Unweighted & 222.73333 & 221.32143 & 0.26994 \\ 
		All\_MaxMatching\_FAMean & 97.83775 & 83.26860 & 0.12360 \\ 
		All\_MaxMatching\_FiberLengthMean & 8061.55115 & 7311.80626 & 0.02059 \\ 
		All\_MaxMatching\_FiberN & 2440.86667 & 2360.00000 & 0.02188 \\ 
		All\_MaxMatching\_FiberNDivLength & 139.27123 & 129.37795 & 0.26989 \\ 
		All\_MaxMatching\_Unweighted & 222.53333 & 220.78571 & 0.16702 \\ 
		All\_MinCutBalDivSum\_FAMean & 0.00992 & 0.01046 & 0.65496 \\ 
		All\_MinCutBalDivSum\_FiberLengthMean & 0.00736 & 0.00859 & 0.34951 \\ 
		All\_MinCutBalDivSum\_FiberN & 0.02266 & 0.02230 & 0.90244 \\ 
		All\_MinCutBalDivSum\_FiberNDivLength & 0.02344 & 0.02198 & 0.61537 \\ 
		All\_MinCutBalDivSum\_Unweighted & 0.00793 & 0.00915 & 0.15562 \\ 
		All\_MinSpanningForest\_FAMean & 96.29849 & 91.93390 & 0.03440 \\ 
		All\_MinSpanningForest\_FiberLengthMean & 5383.39943 & 5298.45575 & 0.02238 \\ 
		All\_MinSpanningForest\_FiberN & 487.20000 & 477.71429 & 0.02138 \\ 
		All\_MinSpanningForest\_FiberNDivLength & 19.13123 & 19.47594 & 0.52402 \\ 
		All\_MinVertexCoverBinary\_Unweighted & 278.33333 & 272.57143 & 0.09750 \\ 
		All\_MinVertexCover\_FAMean & 88.73179 & 83.39334 & 0.00209 & $*$ \\ 
		All\_MinVertexCover\_FiberLengthMean & 8045.57652 & 7329.01371 & 0.02708 \\ 
		All\_MinVertexCover\_FiberN & 2447.73333 & 2363.89286 & 0.01650 \\ 
		All\_MinVertexCover\_FiberNDivLength & 131.52437 & 129.89153 & 0.52678 \\ 
		All\_MinVertexCover\_Unweighted & 222.73333 & 221.32143 & 0.26994 \\ 
		All\_PGEigengap\_FAMean & 0.01011 & 0.01058 & 0.82567 \\ 
		All\_PGEigengap\_FiberLengthMean & 0.00780 & 0.00905 & 0.51801 \\ 
		All\_PGEigengap\_FiberN & 0.01776 & 0.01454 & 0.32360 \\ 
		All\_PGEigengap\_FiberNDivLength & 0.01641 & 0.01372 & 0.37037 \\ 
		All\_PGEigengap\_Unweighted & 0.00888 & 0.00941 & 0.77175 \\ 
		All\_Sum\_FAMean & 996.96869 & 912.06881 & 0.00419 & $*$ \\ 
		All\_Sum\_FiberLengthMean & 72569.22842 & 65893.58253 & 0.02827 \\ 
		All\_Sum\_FiberN & 13388.06667 & 13083.92857 & 0.07726 \\ 
		All\_Sum\_FiberNDivLength & 648.40984 & 652.62526 & 0.75437 \\ 
		All\_Sum\_Unweighted & 2741.13333 & 2671.07143 & 0.19764 \\ 
		Left\_AdjLMaxDivD\_FAMean & 2.13170 & 2.18356 & 0.10520 \\ 
		Left\_AdjLMaxDivD\_FiberLengthMean & 2.28854 & 2.29653 & 0.87385 \\ 
		Left\_AdjLMaxDivD\_FiberN & 4.01054 & 4.39629 & 0.03562 \\ 
		Left\_AdjLMaxDivD\_FiberNDivLength & 3.75976 & 4.06768 & 0.08089 \\ 
		Left\_AdjLMaxDivD\_Unweighted & 1.84695 & 1.85515 & 0.74289 \\ 
		Left\_HoffmanBound\_FAMean & 3.78621 & 3.74262 & 0.52408 \\ 
		Left\_HoffmanBound\_FiberLengthMean & 3.02537 & 2.97013 & 0.43552 \\ 
		Left\_HoffmanBound\_FiberN & 2.51327 & 2.46954 & 0.31630 \\ 
		Left\_HoffmanBound\_FiberNDivLength & 2.49376 & 2.45224 & 0.50396 \\ 
		Left\_HoffmanBound\_Unweighted & 3.86677 & 3.84281 & 0.72661 \\ 
		Left\_LogSpanningForestN\_FAMean & 209.08024 & 184.73327 & 0.00676 & $*$ \\ 
		Left\_LogSpanningForestN\_FiberLengthMean & 1159.44215 & 1136.94931 & 0.06615 \\ 
		Left\_LogSpanningForestN\_FiberN & 727.97774 & 713.31628 & 0.05629 \\ 
		Left\_LogSpanningForestN\_FiberNDivLength & 73.75521 & 68.47575 & 0.40858 \\ 
		Left\_LogSpanningForestN\_Unweighted & 467.56327 & 458.41443 & 0.17067 \\ 
		Left\_MaxFracMatching\_FAMean & 48.19763 & 40.73958 & 0.12892 \\ 
		Left\_MaxFracMatching\_FiberLengthMean & 4050.62536 & 3692.90472 & 0.02154 \\ 
		Left\_MaxFracMatching\_FiberN & 1174.73333 & 1168.14286 & 0.82446 \\ 
		Left\_MaxFracMatching\_FiberNDivLength & 69.30024 & 65.10409 & 0.39737 \\ 
		Left\_MaxFracMatching\_Unweighted & 111.93333 & 111.17857 & 0.27767 \\ 
		Left\_MaxMatching\_FAMean & 48.18911 & 40.61401 & 0.12054 \\ 
		Left\_MaxMatching\_FiberLengthMean & 4081.33975 & 3683.09827 & 0.01115 \\ 
		Left\_MaxMatching\_FiberN & 1171.86667 & 1169.35714 & 0.93445 \\ 
		Left\_MaxMatching\_FiberNDivLength & 69.00646 & 64.86202 & 0.40792 \\ 
		Left\_MaxMatching\_Unweighted & 111.60000 & 110.78571 & 0.24543 \\ 
		Left\_MinCutBalDivSum\_FAMean & 0.09376 & 0.07990 & 0.23745 \\ 
		Left\_MinCutBalDivSum\_FiberLengthMean & 0.08212 & 0.07713 & 0.54194 \\ 
		Left\_MinCutBalDivSum\_FiberN & 0.06769 & 0.06039 & 0.16060 \\ 
		Left\_MinCutBalDivSum\_FiberNDivLength & 0.08034 & 0.05881 & 0.24301 \\ 
		Left\_MinCutBalDivSum\_Unweighted & 0.08572 & 0.07863 & 0.26173 \\ 
		Left\_MinSpanningForest\_FAMean & 46.72246 & 44.99254 & 0.11107 \\ 
		Left\_MinSpanningForest\_FiberLengthMean & 2703.40298 & 2683.84440 & 0.40476 \\ 
		Left\_MinSpanningForest\_FiberN & 245.80000 & 244.35714 & 0.65690 \\ 
		Left\_MinSpanningForest\_FiberNDivLength & 9.60443 & 10.12770 & 0.10870 \\ 
		Left\_MinVertexCoverBinary\_Unweighted & 139.60000 & 136.28571 & 0.06242 \\ 
		Left\_MinVertexCover\_FAMean & 43.55457 & 40.73958 & 0.00274 & $*$ \\ 
		Left\_MinVertexCover\_FiberLengthMean & 4050.47768 & 3692.90472 & 0.02160 \\ 
		Left\_MinVertexCover\_FiberN & 1174.73333 & 1168.14286 & 0.82446 \\ 
		Left\_MinVertexCover\_FiberNDivLength & 65.21177 & 65.10409 & 0.95567 \\ 
		Left\_MinVertexCover\_Unweighted & 111.93333 & 111.17857 & 0.27767 \\ 
		Left\_PGEigengap\_FAMean & 0.07719 & 0.07518 & 0.82459 \\ 
		Left\_PGEigengap\_FiberLengthMean & 0.07879 & 0.07763 & 0.91292 \\ 
		Left\_PGEigengap\_FiberN & 0.06028 & 0.05746 & 0.65896 \\ 
		Left\_PGEigengap\_FiberNDivLength & 0.04513 & 0.04595 & 0.84639 \\ 
		Left\_PGEigengap\_Unweighted & 0.06588 & 0.06319 & 0.71316 \\ 
		Left\_Sum\_FAMean & 488.88692 & 446.37002 & 0.00653 & $*$ \\ 
		Left\_Sum\_FiberLengthMean & 36672.41589 & 33169.82857 & 0.01816 \\ 
		Left\_Sum\_FiberN & 6671.40000 & 6520.07143 & 0.20059 \\ 
		Left\_Sum\_FiberNDivLength & 322.99072 & 325.71679 & 0.73536 \\ 
		Left\_Sum\_Unweighted & 1377.06667 & 1338.42857 & 0.15161 \\ 
		Right\_AdjLMaxDivD\_FAMean & 2.04805 & 2.06086 & 0.77728 \\ 
		Right\_AdjLMaxDivD\_FiberLengthMean & 2.24439 & 2.20712 & 0.61096 \\ 
		Right\_AdjLMaxDivD\_FiberN & 4.21492 & 4.48300 & 0.23523 \\ 
		Right\_AdjLMaxDivD\_FiberNDivLength & 3.69704 & 3.82114 & 0.49290 \\ 
		Right\_AdjLMaxDivD\_Unweighted & 1.81231 & 1.78539 & 0.42677 \\ 
		Right\_HoffmanBound\_FAMean & 3.63851 & 3.57214 & 0.26281 \\ 
		Right\_HoffmanBound\_FiberLengthMean & 2.99319 & 2.98754 & 0.93061 \\ 
		Right\_HoffmanBound\_FiberN & 2.41840 & 2.32451 & 0.02875 \\ 
		Right\_HoffmanBound\_FiberNDivLength & 2.47438 & 2.42756 & 0.32797 \\ 
		Right\_HoffmanBound\_Unweighted & 3.75198 & 3.66399 & 0.14648 \\ 
		Right\_LogSpanningForestN\_FAMean & 218.30058 & 201.22233 & 0.10548 \\ 
		Right\_LogSpanningForestN\_FiberLengthMean & 1144.40045 & 1125.84150 & 0.25579 \\ 
		Right\_LogSpanningForestN\_FiberN & 719.43735 & 708.80001 & 0.25777 \\ 
		Right\_LogSpanningForestN\_FiberNDivLength & 72.66745 & 72.68775 & 0.99730 \\ 
		Right\_LogSpanningForestN\_Unweighted & 461.18676 & 456.60569 & 0.59800 \\ 
		Right\_MaxFracMatching\_FAMean & 49.63129 & 42.48039 & 0.11971 \\ 
		Right\_MaxFracMatching\_FiberLengthMean & 3981.63098 & 3624.50964 & 0.05864 \\ 
		Right\_MaxFracMatching\_FiberN & 1168.00000 & 1134.71429 & 0.16479 \\ 
		Right\_MaxFracMatching\_FiberNDivLength & 67.01619 & 63.26438 & 0.39830 \\ 
		Right\_MaxFracMatching\_Unweighted & 110.76667 & 110.10714 & 0.37162 \\ 
		Right\_MaxMatching\_FAMean & 49.46915 & 42.49032 & 0.12642 \\ 
		Right\_MaxMatching\_FiberLengthMean & 3967.99213 & 3616.44584 & 0.05694 \\ 
		Right\_MaxMatching\_FiberN & 1165.93333 & 1132.21429 & 0.16580 \\ 
		Right\_MaxMatching\_FiberNDivLength & 66.85747 & 63.07703 & 0.39386 \\ 
		Right\_MaxMatching\_Unweighted & 110.60000 & 109.71429 & 0.20386 \\ 
		Right\_MinCutBalDivSum\_FAMean & 0.10473 & 0.08130 & 0.03861 \\ 
		Right\_MinCutBalDivSum\_FiberLengthMean & 0.09399 & 0.07564 & 0.01969 \\ 
		Right\_MinCutBalDivSum\_FiberN & 0.07226 & 0.06309 & 0.03249 \\ 
		Right\_MinCutBalDivSum\_FiberNDivLength & 0.08560 & 0.06297 & 0.20320 \\ 
		Right\_MinCutBalDivSum\_Unweighted & 0.09014 & 0.07466 & 0.00456 & $*$ \\ 
		Right\_MinSpanningForest\_FAMean & 49.66592 & 47.12049 & 0.06306 \\ 
		Right\_MinSpanningForest\_FiberLengthMean & 2674.03630 & 2607.48380 & 0.01248 \\ 
		Right\_MinSpanningForest\_FiberN & 241.86667 & 235.71429 & 0.05077 \\ 
		Right\_MinSpanningForest\_FiberNDivLength & 9.58006 & 9.56983 & 0.97811 \\ 
		Right\_MinVertexCoverBinary\_Unweighted & 138.53333 & 136.07143 & 0.22210 \\ 
		Right\_MinVertexCover\_FAMean & 45.05370 & 42.48039 & 0.01411 \\ 
		Right\_MinVertexCover\_FiberLengthMean & 3981.62212 & 3624.50964 & 0.05865 \\ 
		Right\_MinVertexCover\_FiberN & 1168.00000 & 1134.71429 & 0.16479 \\ 
		Right\_MinVertexCover\_FiberNDivLength & 63.09396 & 63.26438 & 0.92752 \\ 
		Right\_MinVertexCover\_Unweighted & 110.76667 & 110.10714 & 0.37162 \\ 
		Right\_PGEigengap\_FAMean & 0.07888 & 0.05424 & 0.02615 \\ 
		Right\_PGEigengap\_FiberLengthMean & 0.08149 & 0.05411 & 0.02193 \\ 
		Right\_PGEigengap\_FiberN & 0.06247 & 0.04384 & 0.03567 \\ 
		Right\_PGEigengap\_FiberNDivLength & 0.04964 & 0.03607 & 0.05054 \\ 
		Right\_PGEigengap\_Unweighted & 0.06734 & 0.04548 & 0.02094 \\ 
		Right\_Sum\_FAMean & 498.59453 & 455.98757 & 0.02310 \\ 
		Right\_Sum\_FiberLengthMean & 35352.85906 & 32155.12919 & 0.07057 \\ 
		Right\_Sum\_FiberN & 6410.26667 & 6265.28571 & 0.23548 \\ 
		Right\_Sum\_FiberNDivLength & 310.98355 & 312.13891 & 0.87672 \\ 
		Right\_Sum\_Unweighted & 1341.33333 & 1307.64286 & 0.34831 \\ 
	\end{longtable}
	
}

\small{
	\subsection{1015 nodes, round 1}
	\label{Table_Round1_500}
	\begin{longtable}{l | cccc}
		Property & Female & Male & p-value &  \\ 
		All\_AdjLMaxDivD\_FAMean & 3.22321 & 3.30998 & 0.17241 \\ 
		All\_AdjLMaxDivD\_FiberLengthMean & 3.66249 & 3.62589 & 0.72808 \\ 
		All\_AdjLMaxDivD\_FiberN & 9.97079 & 10.49734 & 0.43959 \\ 
		All\_AdjLMaxDivD\_FiberNDivLength & 9.57304 & 10.04031 & 0.47421 \\ 
		All\_AdjLMaxDivD\_Unweighted & 2.78189 & 2.80864 & 0.62854 \\ 
		All\_HoffmanBound\_FAMean & 3.15950 & 3.07640 & 0.01768 \\ 
		All\_HoffmanBound\_FiberLengthMean & 2.72401 & 2.71081 & 0.74127 \\ 
		All\_HoffmanBound\_FiberN & 2.19459 & 2.18310 & 0.77482 \\ 
		All\_HoffmanBound\_FiberNDivLength & 2.19470 & 2.19370 & 0.98348 \\ 
		All\_HoffmanBound\_Unweighted & 3.17194 & 3.14136 & 0.39570 \\ 
		All\_LeftRatio\_FAMean & 0.99025 & 0.98498 & 0.73871 \\ 
		All\_LeftRatio\_FiberLengthMean & 1.02275 & 1.01334 & 0.55556 \\ 
		All\_LeftRatio\_FiberN & 0.99483 & 0.99842 & 0.76208 \\ 
		All\_LeftRatio\_FiberNDivLength & 0.99506 & 0.99977 & 0.65608 \\ 
		All\_LeftRatio\_Unweighted & 1.01401 & 1.00766 & 0.55464 \\ 
		All\_LogSpanningForestN\_FAMean & 439.73425 & 375.35402 & 0.01834 \\ 
		All\_LogSpanningForestN\_FiberLengthMean & 4042.68840 & 3930.97587 & 0.09979 \\ 
		All\_LogSpanningForestN\_FiberN & 2126.38756 & 2075.02830 & 0.10526 \\ 
		All\_LogSpanningForestN\_FiberNDivLength & -361.54957 & -347.10187 & 0.41626 \\ 
		All\_LogSpanningForestN\_Unweighted & 1445.09048 & 1410.64529 & 0.23969 \\ 
		All\_MaxFracMatching\_FAMean & 367.10422 & 392.82651 & 0.48563 \\ 
		All\_MaxFracMatching\_FiberLengthMean & 12444.27073 & 11352.81900 & 0.02610 \\ 
		All\_MaxFracMatching\_FiberN & 2520.13333 & 2450.71429 & 0.02833 \\ 
		All\_MaxFracMatching\_FiberNDivLength & 378.36572 & 407.34233 & 0.46533 \\ 
		All\_MaxFracMatching\_Unweighted & 421.16667 & 412.17857 & 0.11468 \\ 
		All\_MaxMatching\_FAMean & 366.74831 & 392.78175 & 0.47927 \\ 
		All\_MaxMatching\_FiberLengthMean & 12434.60538 & 11353.55886 & 0.02719 \\ 
		All\_MaxMatching\_FiberN & 2518.53333 & 2447.42857 & 0.02355 \\ 
		All\_MaxMatching\_FiberNDivLength & 346.48796 & 404.43920 & 0.23863 \\ 
		All\_MaxMatching\_Unweighted & 420.73333 & 412.21429 & 0.13777 \\ 
		All\_MinCutBalDivSum\_FAMean & 0.01048 & 0.01455 & 0.20028 \\ 
		All\_MinCutBalDivSum\_FiberLengthMean & 0.00513 & 0.00628 & 0.17564 \\ 
		All\_MinCutBalDivSum\_FiberN & 0.02219 & 0.02192 & 0.92698 \\ 
		All\_MinCutBalDivSum\_FiberNDivLength & 0.03288 & 0.03669 & 0.66309 \\ 
		All\_MinCutBalDivSum\_Unweighted & 0.00563 & 0.00672 & 0.06268 \\ 
		All\_MinSpanningForest\_FAMean & 200.77063 & 190.16600 & 0.01194 \\ 
		All\_MinSpanningForest\_FiberLengthMean & 10938.82203 & 10614.65056 & 0.02669 \\ 
		All\_MinSpanningForest\_FiberN & 963.93333 & 939.50000 & 0.01860 \\ 
		All\_MinSpanningForest\_FiberNDivLength & 43.76266 & 43.96543 & 0.83113 \\ 
		All\_MinVertexCoverBinary\_Unweighted & 461.60000 & 448.21429 & 0.10356 \\ 
		All\_MinVertexCover\_FAMean & 152.79728 & 142.97533 & 0.00510 & $*$ \\ 
		All\_MinVertexCover\_FiberLengthMean & 12421.62501 & 11383.69815 & 0.03415 \\ 
		All\_MinVertexCover\_FiberN & 2526.33333 & 2449.85714 & 0.01990 \\ 
		All\_MinVertexCover\_FiberNDivLength & 139.35018 & 137.84455 & 0.55857 \\ 
		All\_MinVertexCover\_Unweighted & 421.96667 & 413.28571 & 0.13796 \\ 
		All\_PGEigengap\_FAMean & 0.00000 & 0.00107 & 0.14466 \\ 
		All\_PGEigengap\_FiberLengthMean & 0.00000 & 0.00103 & 0.16113 \\ 
		All\_PGEigengap\_FiberN & 0.00000 & 0.00190 & 0.13951 \\ 
		All\_PGEigengap\_FiberNDivLength & 0.00000 & 0.00170 & 0.13919 \\ 
		All\_PGEigengap\_Unweighted & 0.00000 & 0.00105 & 0.14705 \\ 
		All\_Sum\_FAMean & 1422.68895 & 1303.27462 & 0.00498 & $*$ \\ 
		All\_Sum\_FiberLengthMean & 99977.79501 & 90954.69035 & 0.03028 \\ 
		All\_Sum\_FiberN & 13586.53333 & 13269.50000 & 0.07020 \\ 
		All\_Sum\_FiberNDivLength & 671.66870 & 674.36023 & 0.84649 \\ 
		All\_Sum\_Unweighted & 3960.20000 & 3831.78571 & 0.14992 \\ 
		Left\_AdjLMaxDivD\_FAMean & 3.15407 & 3.26128 & 0.06781 \\ 
		Left\_AdjLMaxDivD\_FiberLengthMean & 3.48214 & 3.50359 & 0.79761 \\ 
		Left\_AdjLMaxDivD\_FiberN & 7.25286 & 8.08325 & 0.02928 \\ 
		Left\_AdjLMaxDivD\_FiberNDivLength & 6.94687 & 7.51631 & 0.14162 \\ 
		Left\_AdjLMaxDivD\_Unweighted & 2.69638 & 2.74517 & 0.33713 \\ 
		Left\_HoffmanBound\_FAMean & 3.22364 & 3.18000 & 0.37811 \\ 
		Left\_HoffmanBound\_FiberLengthMean & 2.75210 & 2.71714 & 0.43501 \\ 
		Left\_HoffmanBound\_FiberN & 2.40776 & 2.36099 & 0.30309 \\ 
		Left\_HoffmanBound\_FiberNDivLength & 2.38876 & 2.37013 & 0.70703 \\ 
		Left\_HoffmanBound\_Unweighted & 3.23817 & 3.19290 & 0.33180 \\ 
		Left\_LogSpanningForestN\_FAMean & 210.85942 & 178.12745 & 0.02486 \\ 
		Left\_LogSpanningForestN\_FiberLengthMean & 2032.22116 & 1971.64825 & 0.07003 \\ 
		Left\_LogSpanningForestN\_FiberN & 1068.89981 & 1041.23415 & 0.09149 \\ 
		Left\_LogSpanningForestN\_FiberNDivLength & -179.63339 & -171.63161 & 0.51982 \\ 
		Left\_LogSpanningForestN\_Unweighted & 728.12558 & 708.39172 & 0.18427 \\ 
		Left\_MaxFracMatching\_FAMean & 183.97769 & 197.14160 & 0.47430 \\ 
		Left\_MaxFracMatching\_FiberLengthMean & 6296.31570 & 5736.50436 & 0.01837 \\ 
		Left\_MaxFracMatching\_FiberN & 1220.00000 & 1213.53571 & 0.81314 \\ 
		Left\_MaxFracMatching\_FiberNDivLength & 188.25111 & 203.66306 & 0.43945 \\ 
		Left\_MaxFracMatching\_Unweighted & 211.53333 & 206.89286 & 0.13059 \\ 
		Left\_MaxMatching\_FAMean & 183.69446 & 196.98025 & 0.46852 \\ 
		Left\_MaxMatching\_FiberLengthMean & 6290.30035 & 5734.95255 & 0.01894 \\ 
		Left\_MaxMatching\_FiberN & 1218.86667 & 1214.35714 & 0.86884 \\ 
		Left\_MaxMatching\_FiberNDivLength & 188.08013 & 203.30161 & 0.44482 \\ 
		Left\_MaxMatching\_Unweighted & 211.20000 & 206.64286 & 0.14188 \\ 
		Left\_MinCutBalDivSum\_FAMean & 0.09885 & 0.10926 & 0.69063 \\ 
		Left\_MinCutBalDivSum\_FiberLengthMean & 0.05144 & 0.04757 & 0.51458 \\ 
		Left\_MinCutBalDivSum\_FiberN & 0.04602 & 0.04116 & 0.20617 \\ 
		Left\_MinCutBalDivSum\_FiberNDivLength & 0.20876 & 0.20881 & 0.99933 \\ 
		Left\_MinCutBalDivSum\_Unweighted & 0.05227 & 0.04713 & 0.21386 \\ 
		Left\_MinSpanningForest\_FAMean & 97.33236 & 92.63020 & 0.06143 \\ 
		Left\_MinSpanningForest\_FiberLengthMean & 5481.16240 & 5324.53691 & 0.04038 \\ 
		Left\_MinSpanningForest\_FiberN & 483.26667 & 473.42857 & 0.14910 \\ 
		Left\_MinSpanningForest\_FiberNDivLength & 21.95206 & 22.26390 & 0.52551 \\ 
		Left\_MinVertexCoverBinary\_Unweighted & 232.33333 & 225.28571 & 0.10878 \\ 
		Left\_MinVertexCover\_FAMean & 75.44749 & 70.49228 & 0.00733 & $*$ \\ 
		Left\_MinVertexCover\_FiberLengthMean & 6259.65401 & 5759.90764 & 0.03606 \\ 
		Left\_MinVertexCover\_FiberN & 1218.16667 & 1210.60714 & 0.77425 \\ 
		Left\_MinVertexCover\_FiberNDivLength & 69.27402 & 69.53888 & 0.88850 \\ 
		Left\_MinVertexCover\_Unweighted & 211.73333 & 207.50000 & 0.17616 \\ 
		Left\_PGEigengap\_FAMean & 0.01178 & 0.01763 & 0.54919 \\ 
		Left\_PGEigengap\_FiberLengthMean & 0.01241 & 0.01922 & 0.52171 \\ 
		Left\_PGEigengap\_FiberN & 0.00979 & 0.01642 & 0.43541 \\ 
		Left\_PGEigengap\_FiberNDivLength & 0.00689 & 0.01269 & 0.34893 \\ 
		Left\_PGEigengap\_Unweighted & 0.00976 & 0.01522 & 0.50417 \\ 
		Left\_Sum\_FAMean & 703.63345 & 641.32910 & 0.00531 & $*$ \\ 
		Left\_Sum\_FiberLengthMean & 50902.69786 & 46025.48488 & 0.01756 \\ 
		Left\_Sum\_FiberN & 6766.13333 & 6615.71429 & 0.20853 \\ 
		Left\_Sum\_FiberNDivLength & 334.51749 & 336.88634 & 0.77548 \\ 
		Left\_Sum\_Unweighted & 2004.66667 & 1929.57143 & 0.10019 \\ 
		Right\_AdjLMaxDivD\_FAMean & 3.11058 & 3.17176 & 0.44453 \\ 
		Right\_AdjLMaxDivD\_FiberLengthMean & 3.48779 & 3.51830 & 0.79334 \\ 
		Right\_AdjLMaxDivD\_FiberN & 7.76024 & 8.14382 & 0.37225 \\ 
		Right\_AdjLMaxDivD\_FiberNDivLength & 6.83772 & 7.07156 & 0.52127 \\ 
		Right\_AdjLMaxDivD\_Unweighted & 2.70744 & 2.71957 & 0.83883 \\ 
		Right\_HoffmanBound\_FAMean & 3.14757 & 3.04963 & 0.02036 \\ 
		Right\_HoffmanBound\_FiberLengthMean & 2.77024 & 2.68725 & 0.05471 \\ 
		Right\_HoffmanBound\_FiberN & 2.32441 & 2.23330 & 0.01858 \\ 
		Right\_HoffmanBound\_FiberNDivLength & 2.38419 & 2.30535 & 0.07109 \\ 
		Right\_HoffmanBound\_Unweighted & 3.16199 & 3.09316 & 0.07931 \\ 
		Right\_LogSpanningForestN\_FAMean & 223.08908 & 190.45255 & 0.05945 \\ 
		Right\_LogSpanningForestN\_FiberLengthMean & 2000.50088 & 1948.56703 & 0.17197 \\ 
		Right\_LogSpanningForestN\_FiberN & 1048.39181 & 1024.37929 & 0.19504 \\ 
		Right\_LogSpanningForestN\_FiberNDivLength & -187.57760 & -181.71216 & 0.49112 \\ 
		Right\_LogSpanningForestN\_Unweighted & 710.28043 & 694.54365 & 0.34901 \\ 
		Right\_MaxFracMatching\_FAMean & 182.91341 & 195.61348 & 0.49409 \\ 
		Right\_MaxFracMatching\_FiberLengthMean & 6136.77547 & 5605.35278 & 0.05092 \\ 
		Right\_MaxFracMatching\_FiberN & 1194.00000 & 1169.53571 & 0.23073 \\ 
		Right\_MaxFracMatching\_FiberNDivLength & 186.50824 & 201.72223 & 0.44500 \\ 
		Right\_MaxFracMatching\_Unweighted & 209.40000 & 205.17857 & 0.18566 \\ 
		Right\_MaxMatching\_FAMean & 182.80549 & 195.43970 & 0.49551 \\ 
		Right\_MaxMatching\_FiberLengthMean & 6133.19214 & 5607.85845 & 0.05338 \\ 
		Right\_MaxMatching\_FiberN & 1193.33333 & 1168.92857 & 0.23115 \\ 
		Right\_MaxMatching\_FiberNDivLength & 186.33486 & 201.44535 & 0.44822 \\ 
		Right\_MaxMatching\_Unweighted & 209.20000 & 205.21429 & 0.21712 \\ 
		Right\_MinCutBalDivSum\_FAMean & 0.10194 & 0.10852 & 0.79465 \\ 
		Right\_MinCutBalDivSum\_FiberLengthMean & 0.05798 & 0.05045 & 0.10536 \\ 
		Right\_MinCutBalDivSum\_FiberN & 0.04848 & 0.04187 & 0.01462 \\ 
		Right\_MinCutBalDivSum\_FiberNDivLength & 0.22021 & 0.21702 & 0.95938 \\ 
		Right\_MinCutBalDivSum\_Unweighted & 0.05808 & 0.04940 & 0.01406 \\ 
		Right\_MinSpanningForest\_FAMean & 103.51247 & 97.80199 & 0.02535 \\ 
		Right\_MinSpanningForest\_FiberLengthMean & 5449.19539 & 5284.24222 & 0.05427 \\ 
		Right\_MinSpanningForest\_FiberN & 482.53333 & 470.00000 & 0.08846 \\ 
		Right\_MinSpanningForest\_FiberNDivLength & 21.91228 & 22.07628 & 0.81623 \\ 
		Right\_MinVertexCoverBinary\_Unweighted & 229.06667 & 222.71429 & 0.16269 \\ 
		Right\_MinVertexCover\_FAMean & 77.19144 & 72.32630 & 0.01652 \\ 
		Right\_MinVertexCover\_FiberLengthMean & 6150.67656 & 5612.96934 & 0.04968 \\ 
		Right\_MinVertexCover\_FiberN & 1200.03333 & 1171.71429 & 0.16130 \\ 
		Right\_MinVertexCover\_FiberNDivLength & 66.59407 & 66.55426 & 0.97982 \\ 
		Right\_MinVertexCover\_Unweighted & 209.93333 & 205.64286 & 0.18522 \\ 
		Right\_PGEigengap\_FAMean & 0.01467 & 0.00967 & 0.56364 \\ 
		Right\_PGEigengap\_FiberLengthMean & 0.01533 & 0.00998 & 0.55712 \\ 
		Right\_PGEigengap\_FiberN & 0.01218 & 0.00779 & 0.54063 \\ 
		Right\_PGEigengap\_FiberNDivLength & 0.00960 & 0.00631 & 0.56034 \\ 
		Right\_PGEigengap\_Unweighted & 0.01225 & 0.00787 & 0.54125 \\ 
		Right\_Sum\_FAMean & 709.28257 & 651.61088 & 0.02201 \\ 
		Right\_Sum\_FiberLengthMean & 48529.39175 & 44331.93496 & 0.07380 \\ 
		Right\_Sum\_FiberN & 6514.80000 & 6355.14286 & 0.19297 \\ 
		Right\_Sum\_FiberNDivLength & 322.73850 & 322.68174 & 0.99406 \\ 
		Right\_Sum\_Unweighted & 1932.06667 & 1875.71429 & 0.28274 \\ 
	\end{longtable}
	
}

\end{document}